\documentclass[preprint, 5p]{elsarticle}

\usepackage{xcolor}
\usepackage{lineno}
\usepackage{url}
\usepackage{booktabs}
\usepackage{graphicx}
\usepackage{caption}
\usepackage{subcaption}
\usepackage{amsmath}
\usepackage{float}
\usepackage{imakeidx}
\usepackage[utf8]{inputenc}

\usepackage{natbib}
\journal{Nuclear Inst. and Methods in Physics Research, A}

\begin{document}

\begin{frontmatter}
\title{A parametric approach for the identification of single-charged isotopes with AMS-02}

\author[1]{E. F. Bueno\corref{cor1}}
\ead{e.ferronato.bueno@rug.nl}
\author[2]{F. Barão}
\ead{barao@lip.pt}
\author[1]{M. Vecchi}
\ead{m.vecchi@rug.nl}

\cortext[cor1]{Corresponding author}

\address[1]{Kapteyn Astronomical Institute, University of Groningen,Landleven 12, 9747 AD, Groningen, the Netherlands}
\address[2]{Laboratório de Instrumentação e Física Experimental de Partículas (LIP), 1649-003 Lisboa, Portugal}

\begin{abstract}

Measurements of the isotopic composition of single-charged cosmic rays provide important insights in the propagation processes. However, the isotopic identification is challenging due to the one hundred times greater abundance of protons when compared to deuterons, the only stable isotope of hydrogen. Taking advantage of the precise measurements of the velocity and momentum in the Alpha Magnetic Spectrometer (AMS-02), a particle physics detector operating aboard the International Space Station since May 2011, we describe a parametric template fit method, which takes into account systematic uncertainties such as the fragmentation of particles inside AMS-02 and eventual differences between data and simulation through the use of nuisance parameters. With this method we are also able to assess the AMS-02 performance in terms of mass resolution, showing that it is able to separate the isotopes of hydrogen up to 10 GeV/n.

\end{abstract}

\end{frontmatter}

\section{Introduction}
\label{intro}

It is well known that the cosmic-ray (CR) spectrum is dominated by positive, single-charged nuclei, with almost the entirety of such nuclei being protons\citep{gaisser}. These particles can be primary, when they are produced and accelerated at the CR sources, and secondary, when they are produced during the interactions of primary particles with the interstellar medium (ISM) during their propagation \citep{coste}. The isotopic composition of hydrogen in cosmic-rays is dominated by protons, but there is also a small component of deuterons, which are also stable. Deuterons are produced in the first step of the proton-proton chain in stars, but they are readily consumed in the next step \citep{ppchain}. Hence, the deuterons in CRs are expected to be mostly of secondary origin. That is, they are produced due to the interactions of certain CRs, mainly p, $^{3}$He, and $^{4}$He, with the ISM. Therefore, the measurement of the deuteron flux and the corresponding secondary-to-primary ratios, such as the deuteron-to-proton (d/p) and deuteron-to-helium-4 ($\mbox{d}/^{4}\mbox{He}$) flux ratios provide valuable insights into the CR propagation processes which take place in our galaxy \citep{coste}.

The isotopic composition of single-charged CRs has been studied by several experiments, with some of its first measurements being performed in the sixties by experiments such as IMP-3\citep{IMP3}, which showed that the d/p ratio is as low as $5\%$ at a few tens of MeV/nucleon. In more recent times, experiments such as PAMELA\citep{PAMELA}, BESS\citep{BESS00}, CAPRICE\citep{CAPRICE98}, and IMAX\citep{IMAX92} have extended the energy range of the measurements, showing that deuterons are of the order of $1\%$ of the protons up to 4 GeV/n. Several methods have been applied to perform the isotopic identification, by combining the rigidity, $R = pc/Ze$, and velocity, $\beta$ (or equivalently $dE/dX$, which is proportional to $ \sim \beta^{-2}$), to use the mass difference between the isotopes to separate them, either by using cut-based methods or template fits in one of the variables.  The mass as a function of the velocity and rigidity is given by the equation

\begin{equation}
    m = \frac{RZ}{\beta\gamma}
\label{eq:mass}
\end{equation}
\noindent where $R$ is the rigidity, $Z$ is the magnitude of the charge; $\beta$ is the velocity in speed-of-light units, and  $\gamma$ is the Lorentz factor.

Considering that the proton background is about one hundred times larger than the deuterons, the mass resolution is what ultimately determines the limits of these measurements. From equation \ref{eq:mass}, the the derivation of the mass resolution gives

\begin{equation}
    \left(\frac{\Delta m}{m}\right)^{2} =\left(\frac{\Delta R}{R}\right)^{2} + \gamma^{4}\left(\frac{\Delta \beta}{\beta}\right)^{2}
\label{eq:massres}
\end{equation}

\noindent The dependence on the fourth power of the Lorentz factor indicates that the mass resolution increases rapidly with $\beta$. This effect was seen in the mass resolution reported by the PAMELA experiment, where they obtained values between $6.5$ and $10\%$ for helium isotopes (which have better resolution values than single-charged isotopes due to the strength of the signal) in the rigidity range between 1 and 3.5 GV \citep{PAMELA}, rapidly increasing with the rigidity as expected from equation \ref{eq:massres}. Hence, the identification of the isotopes through their mass at higher energies requires an excellent velocity resolution.

\section{The Alpha Magnetic Spectrometer}

\begin{figure}[!h]
  \centering{\includegraphics[scale=0.6]{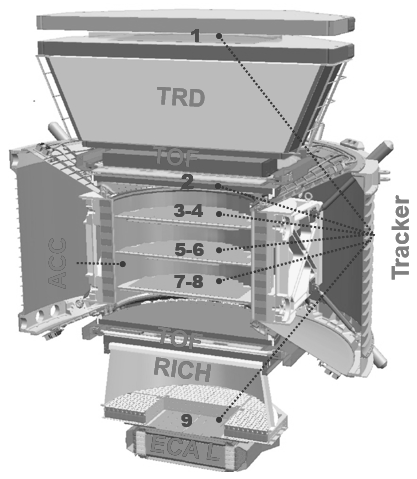}}
  \caption{Schematic view of the Alpha Magnetic Spectrometer (AMS-02) detector scheme. See text for discussion. Figure taken from reference \citep{amsscheme}.
    \label{fig:ams} 
  }
\end{figure}

The Alpha Magnetic Spectrometer (AMS-02), shown in figure \ref{fig:ams}, a particle physics detector operating aboard the International Space Station (ISS) since May 2011, can perform the isotopic identification in a broader energy range than previous experiments because of the precise and complementary velocity measurements of its Time of Flight and Cherenkov detectors. AMS-02 is composed of several sub-detectors \citep{AMSPhysReport}: the nine layers of the silicon tracker, together with the 0.15 T permanent magnet, measure the charge, charge sign, and the rigidity ($R = pc/Ze$) of the particle; the Transition Radiator Detector (TRD) is used to separate leptons from hadrons; four Time of Flight (TOF) scintillator planes are used to measure the charge and velocity of the particle and act as the main trigger of the experiment; the Anti-Coincidence Counter (ACC) is used to reject particles with high incidence angle; the Ring Imaging Cherenkov Detector (RICH) provides a very precise velocity measurement and also measures the electric charge; finally, the Electromagnetic Calorimeter (ECAL) is used for lepton-hadron separation and the measurement of the energy of the particle.

The Time of Flight (ToF) detector is made of four layers of scintillator planes, placed in pairs above (Upper-ToF) and below (Lower-ToF) the magnet. When a particle travels on a path through the magnetic field between the scintillators, the time taken, $\Delta t$, is registered; the particle's trajectory length in the magnetic field, $L$, is obtained with the reconstructed tracker information. Given the time and distance, the velocity, $\beta$, is simply given by  $\beta = \Delta t/cL$, and is measured with a resolution $\Delta \beta / \beta^{2} \approx 4\% $ for particles with $Z=1$ and $\beta \approx 1$ \citep{TOFPerformance}.

The RICH, a proximity focusing detector, measures the velocity through the Cherenkov effect: particles crossing the radiator plane emit electromagnetic radiation peaked on the UV region\citep{amsrich}. Below this radiator, there is a detection plane, made of $4 \times 4$ pixelized photomultipliers, where the photons are collected. In order to reduce the amount of matter traversed by particles before the ECAL, the PMTs of the detection plane are arranged to be outside the calorimeter acceptance. Using the coordinates of the hits, the Cherenkov ring is reconstructed, enabling the calculation of the velocity through the Cherenkov angle. The  RICH is made of two radiator materials; in the center sodium fluoride (NaF), with a refraction index $n = 1.33$ and threshold of emission $\beta = 0.75$, while surrounding it exists the Aerogel (AGL), with a refraction index $n = 1.05$ and threshold of emission $\beta = 0.96$. The velocity resolution in the NaF is $\Delta \beta / \beta \approx 0.35\% $ for particles with $Z=1$ and $\beta \approx 1$; in the same conditions the AGL has a resolution of  $\Delta \beta / \beta \approx 0.12\%$ \citep{AMSPhysReport}.

The tracker \citep{AMSPhysReport} consists of 9 silicon layers placed along the body of the detector, with layer 1 being located on top of the TRD and layer 9 right above the ECAL. In contrast, layers from 2 to 8 are located within the magnet, constituting the inner tracker. Each of these nine layers registers the two-dimensional coordinates of the traversing particle, enabling the reconstruction of the trajectory and, therefore, the curvature of the particles inside the magnetic field, inversely proportional to the rigidity. The resolution of the tracker is $\Delta R/R \approx 0.1$ for particles with $R < 20 $ GV \citep{AMSPhysReport}.

In this work, 8.5 years of data collected by AMS, from May 2011 to January 2020, were used. Events compatible with $Z = 1$ were selected in different parts of the detector by using the charge measured by the inner tracker, upper and lower ToF. Furthermore, to reduce the $Z \geq 2$ fragmentation background from fragmentation at the top of the detector, events have their charge selected at the first layer of the tracker, $Z_{L1},$ to be compatible with one. For this reason, a cut is applied on the charge calculated using the highest available signal on L1. Since the highest signal is used, this charge distribution is positively skewed, hence an asymmetric charge window is selected, namely $0.8 < Z_{L1} < 1.6$. Events were required to have hits on the four planes of the ToF to ensure a good velocity reconstruction. The events were required to have Cherenkov photons detected by at least three different PMTs of the RICH, with at least $50\%$ of the hits detected by the PMT plane are being used on the ring reconstruction, which ensure that noisy events are rejected. Additionally, to ensure the quality of the track reconstruction, events were required to have $\chi^{2}/NDF$ of the track fitting procedure below 10 in both the bending and non-bending coordinates.

Given that the $\beta$ resolution is in the range of $10^{-2}$ to $10^{-3}$  in the velocity interval of the analysis, the events were binned in this variable, with overlapping ranges. The ToF was used for events with $0.5 < \beta < 0.9$, the NaF for events with $0.75 < \beta < 0.98$  and the AGL for events  with $0.96 < \beta < 0.997$. In the case of the ToF, the lower limit was determined by the minimum velocity for the particle to cross the magnetic field, whereas, in the RICH, it was determined by the emission threshold for Cherenkov photons. The velocity resolution of each detector determined the upper limit for the relevant velocity range. To select only CRs of galactic origin, a geomagnetic cutoff selection is also applied. Given that the binning of the data is in velocity, the selection is also applied with this variable: events are selected only if $\beta > \beta_{C}(m_{i}, R_{C}$), where $\beta_{C}$ is the geomagnetic cutoff velocity calculated using $R_{C}$, the cutoff rigidity calculated with the most recent IGRF geomagnetic model \citep{igrf}, and $m_{i}$ is the mass of the isotope of interest. For this study, we have used the deuteron mass as input because as deuterons have a higher mass, the cutoff velocity is lower, increasing the number of events of the deuteron sample as compared to what we would have if the proton mass was used as input.

In addition to the experimental data, Monte Carlo (MC) simulations of protons and deuterons were also used in this study. These simulations were produced by the AMS collaboration through a dedicated software based on the GEANT4 package \cite{GEANT4}. The software simulates the interactions of particles with AMS material and produces detector responses, which are then used to reconstruct the desired event properties the same way as in data.

\section{Mass identification in AMS-02}

The isotope identification in this analysis is performed via the mass measurement by combining the rigidity measured by the tracker and the velocity measured by either the ToF or the RICH according to equation \ref{eq:mass}. In this section, the methods for extracting the signals are presented and discussed.

\subsection{Methodology}
 The event counting can be done with different techniques, either cut-based analyses or template fits. The latter has been chosen for this work, as the overlapping mass regions are automatically dealt with during the fit procedure. In addition, parametric templates were chosen over non-parametric ones for two main reasons: the parametrization is continuous, therefore regularizing the data, avoiding fluctuations, and the estimation of the systematic uncertainties can be done quickly through the usage of nuisance parameters.
Considering the importance of the rigidity for the shape of the mass and the fact that the tracker measures the sagitta \citep{AMSPhysReport}, which is proportional to the inverse of the rigidity, $1/R$, the templates are defined in terms of inverse mass. As a result, such distributions can be parametrized by combining a Gaussian and slightly modified Gaussians. As velocity comes from measurements relying upon two different techniques (ToF and RICH), the mass parametrizations are slightly different to accommodate the different backgrounds on each reconstruction technique, which will be presented and discussed further in this paper. The model parameters were studied as a function of $\beta$ with MC samples, and their behavior was regularized using cubic splines.

In the following section, we will describe the templates of both protons and deuterons and how their parameters change with velocity.

\subsection{Proton templates}

\begin{figure}[!h]
  \centering{\includegraphics[width=\linewidth]{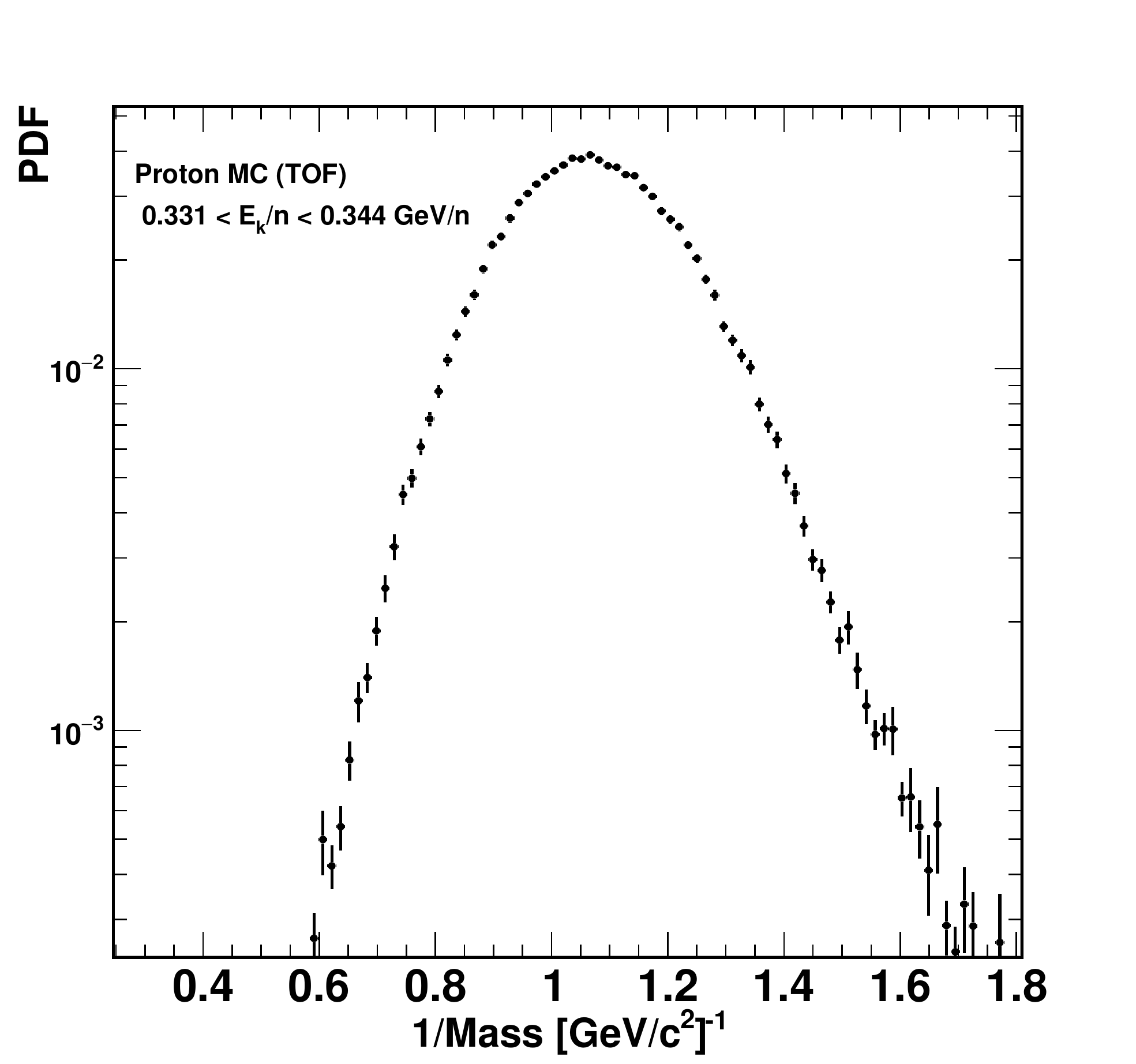}}
  \caption{Example of the normalized inverse mass distribution of proton MC in the ToF range for events with kinetic energy per nucleon between 0.331 and 0.344 GeV/n.
    \label{fig:mass_tof} 
  }
\end{figure}

As discussed in section \ref{intro}, in the velocity interval \\ $0.5 < \beta < 0.9$ the measurement of the velocity is done with the ToF. In this detector, the main source of background comes from the interactions and multiple-scattering in AMS, which produce a low mass (high inverse-mass) tail on the distribution, as seen in figure \ref{fig:mass_tof}, where a typical example of the ToF inverse mass distribution is shown for simulated proton events detected by AMS.

In this range, in each velocity bin, the inverse mass distribution
 is parametrized by a Gaussian core ($G$) summed to an exponentially-modified Gaussian ($EMG)$ with an exponential tail to the right (to describe the high-inverse mass events), that is:

\begin{multline}
M_{p}(1/m) = f_{G} \cdot  G (1/m; \mu, \sigma)\\
 + (1 - f_{G} ) \cdot EMG( 1/m; \mu, \xi\sigma, \lambda_{1} )
\end{multline}
\noindent where $\mu$ and $\sigma$ indicate respectively the mean and the width of the Gaussian, $f_{G}$ is the fraction of the Gaussian contribution, $\lambda_{1}$ is the tail parameter, and $\xi$ the factor which determines the increase of the EMG width with respect to the Gaussian. The parameter $f_{G}$ is written as a function of $\mu$, $\lambda_{1}$ and the expected value of the inverse mass distribution, $\langle1/m\rangle$. The expected value of the parametric model is given by the weighted sum of two components: a Gaussian component, $\mu$, and a component from the EMG, which has its expected value given by $\mu + 1/\lambda_{1}$, which leads to:

\begin{equation}
    \langle1/m\rangle = f_{G}\cdot\mu + (1-f_{G})\cdot(\mu + 1/\lambda_{1})
\end{equation}

\noindent therefore 

\begin{equation}
    f_{G} = 1 + \lambda_{1} \cdot(\mu - \langle1/m\rangle)
\end{equation}

\noindent where $\langle1/m\rangle$ is computed numerically from the histogram of the inverse mass distribution.

\begin{figure}[!h]
  \centering{\includegraphics[width=\linewidth]{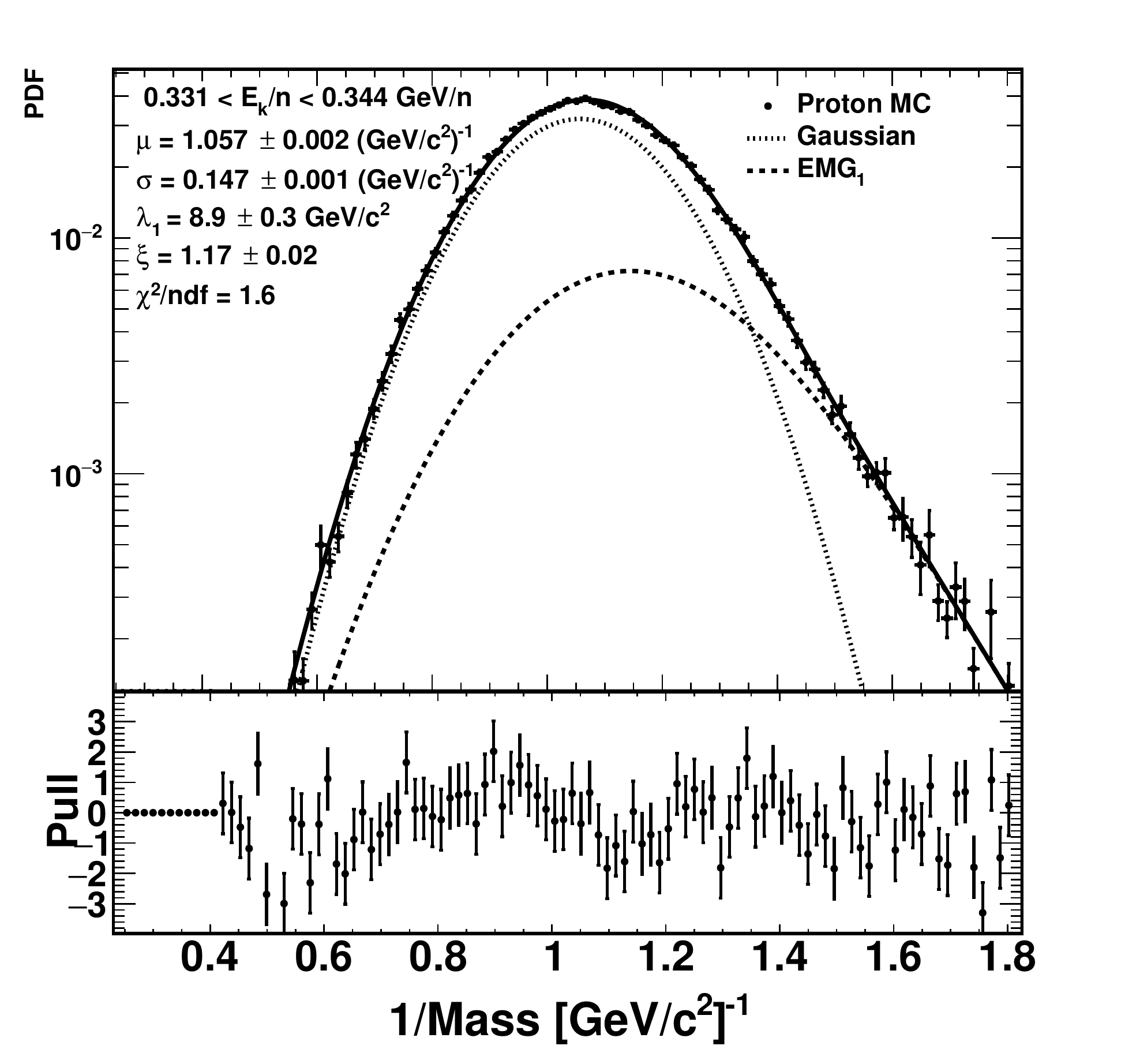}}
  \caption{Example of fit of the parametric model to proton MC in the ToF range for events with kinetic energy per nucleon between 0.331 and 0.344 GeV/n (top panel). The different lines types indicate the components of the model: the Gaussian is the thin-dashed line;  the EMG is the thick-dashed line, while their sum is in solid black. The bottom panel shows the pull of the fit.
    \label{fig:fit_mc_tof} 
}
\end{figure}

 A typical example of fit for simulated proton events is shown in figure \ref{fig:fit_mc_tof}, with $\chi^{2}/\text{ndf} = 1.6$. It is clear from the figure that the proposed model describes the MC quite well, which is also stressed by the pull\footnote{The pull is defined as the difference between the data, $x$, and the model, $\theta$, normalized by the total uncertainty (data + model), $\sigma$. That is $p = \frac{x-\theta}{\sigma}$.}.

\begin{figure}[!h]
\centering{\includegraphics[width=\linewidth]{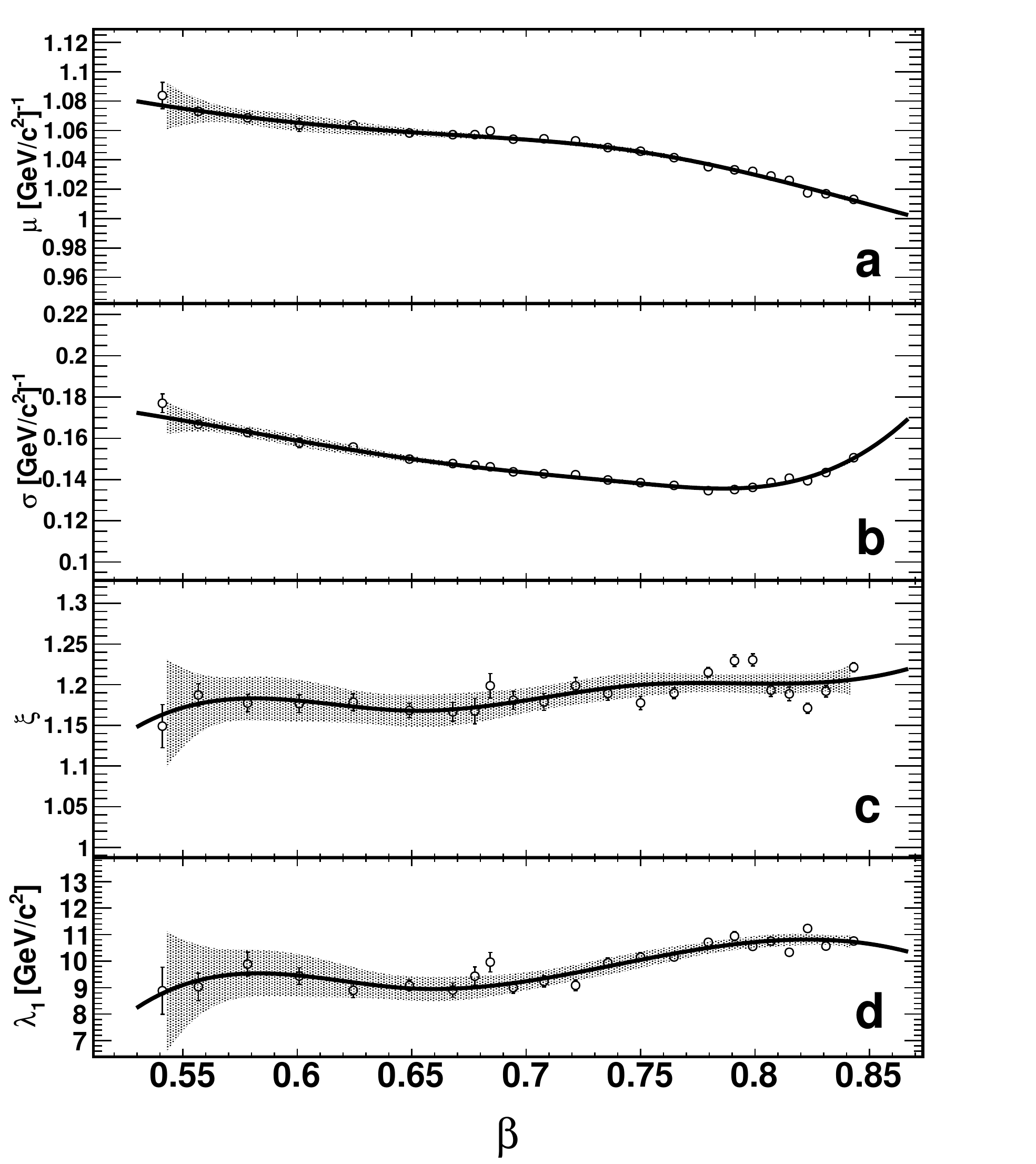}}
\caption{The parameters of the model as a function of $\beta$ in the ToF range. Panel (a) shows $\mu$, the mean of the gaussian; (b) shows $\sigma$, its width; panel (c) shows $\xi$, the scaling factor which determines the increase of the EMG width with respect to the Gaussian; and panel (d) shows $\lambda_{1}$, the exponential parameter of the EMG. The black curve represents the spline fit in all panels, and the gray bands the 95$\%$ confidence intervals.
    \label{fig:pars_tof}
}
\end{figure}

The velocity dependence of the other parameters, namely $\mu$, $\sigma$, $\xi$ and $\lambda_{1}$, are shown in figure \ref{fig:pars_tof}. The behavior of $\mu$ reflects the fact in the ToF, the energy loss suffered by the particles shifts the inverse mass peak to the right (reducing the mass). As the velocity increases, the value gets closer to the expected proton mass, $1/m = 1.06 (\text{GeV/c}^{2})^{-1}$ , as shown in panel (a). The parameter $\sigma$, shown in panel (b),  represents the width of the inverse mass distribution. One can note that at first, its value decreases from 0.18 until around 0.14 and then, at $\beta > 0.8$ starts to increase again. These two behaviors come from different physical effects: the rigidity resolution decreases at low energy due to multiple scattering and progressively gets better, reaching a plateau; then, the $\gamma^{2}$ factor of the mass resolution ($\Delta m/m$) becomes dominant as the velocity increases, increasing $\sigma$. The parameter $\xi$, shown in panel (c), which describes the width of the EMG with respect to the Gaussian, shows the that EMG has a width approximately 15$\%$ larger in the entire range. 
The behavior of the tail parameter, $\lambda_{1}$ is shown in panel (d). Since it represents the exponential mass tail, the lower values of $\lambda_{1}$ at lower velocities indicate the tail is more relevant in these region, becoming slightly less important as the velocity increases.

\begin{figure}[!h]
  \centering{\includegraphics[width=\linewidth]{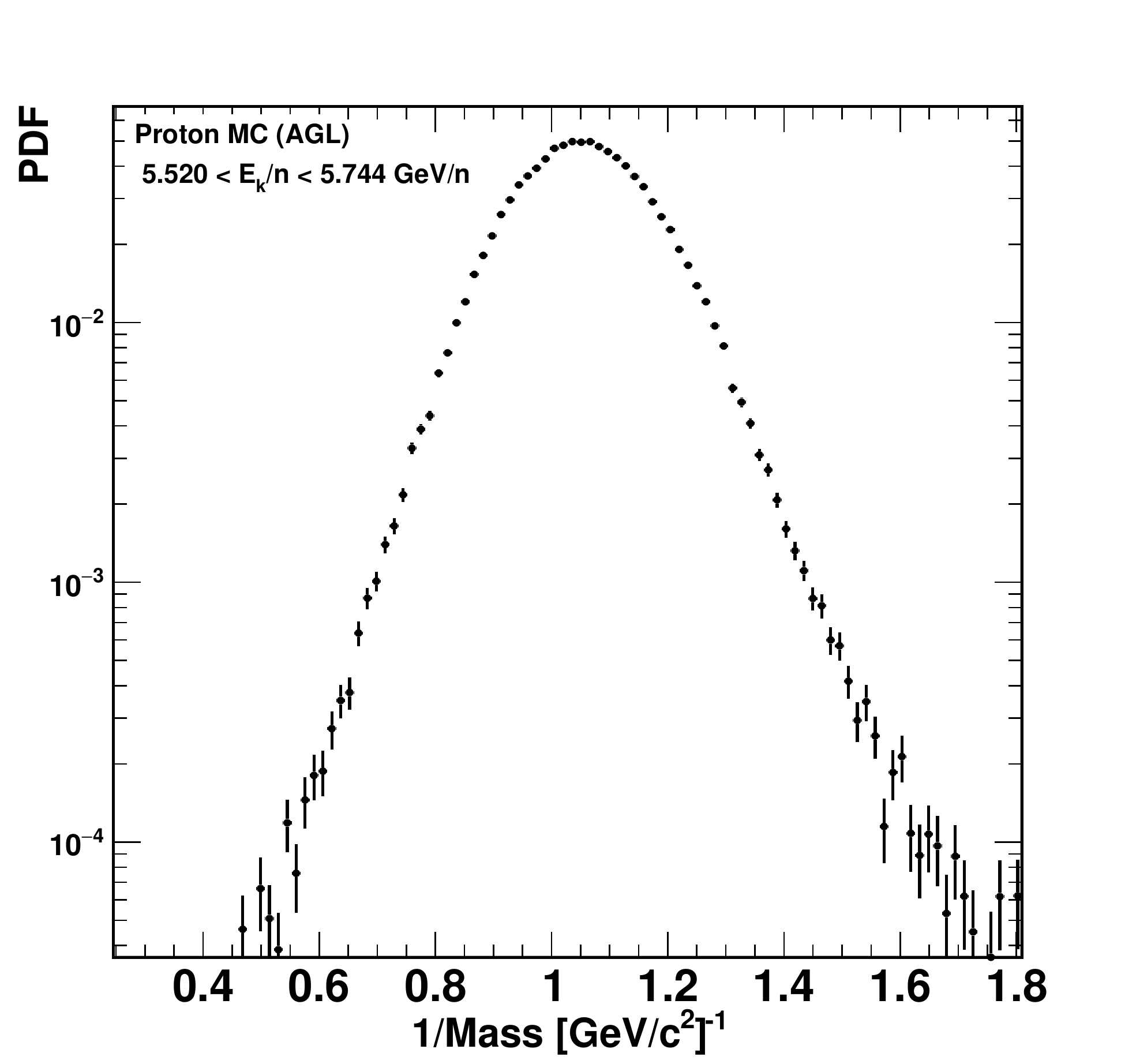}}
\caption{Example of the inverse mass distribution of proton MC in the RICH-Aerogel region for events with kinetic energy per nucleon between 5.520 and 5.744 GeV/n.
  \label{fig:mass_rich}
}
\end{figure}

The shape of the inverse mass distributions obtained using the velocity measured by the RICH (including both NaF and Aerogel) typically consists of a Gaussian core which is then enlarged by instrumental effects. As in the ToF, a low-mass tail comes from interactions with detector material during the propagation of the incident cosmic ray inside the detector. However, a tail of high mass events is also present, as shown in figure \ref{fig:mass_rich}, explained by interactions inside AMS.

\begin{figure}[!h]
  \centering{\includegraphics[width=\linewidth]{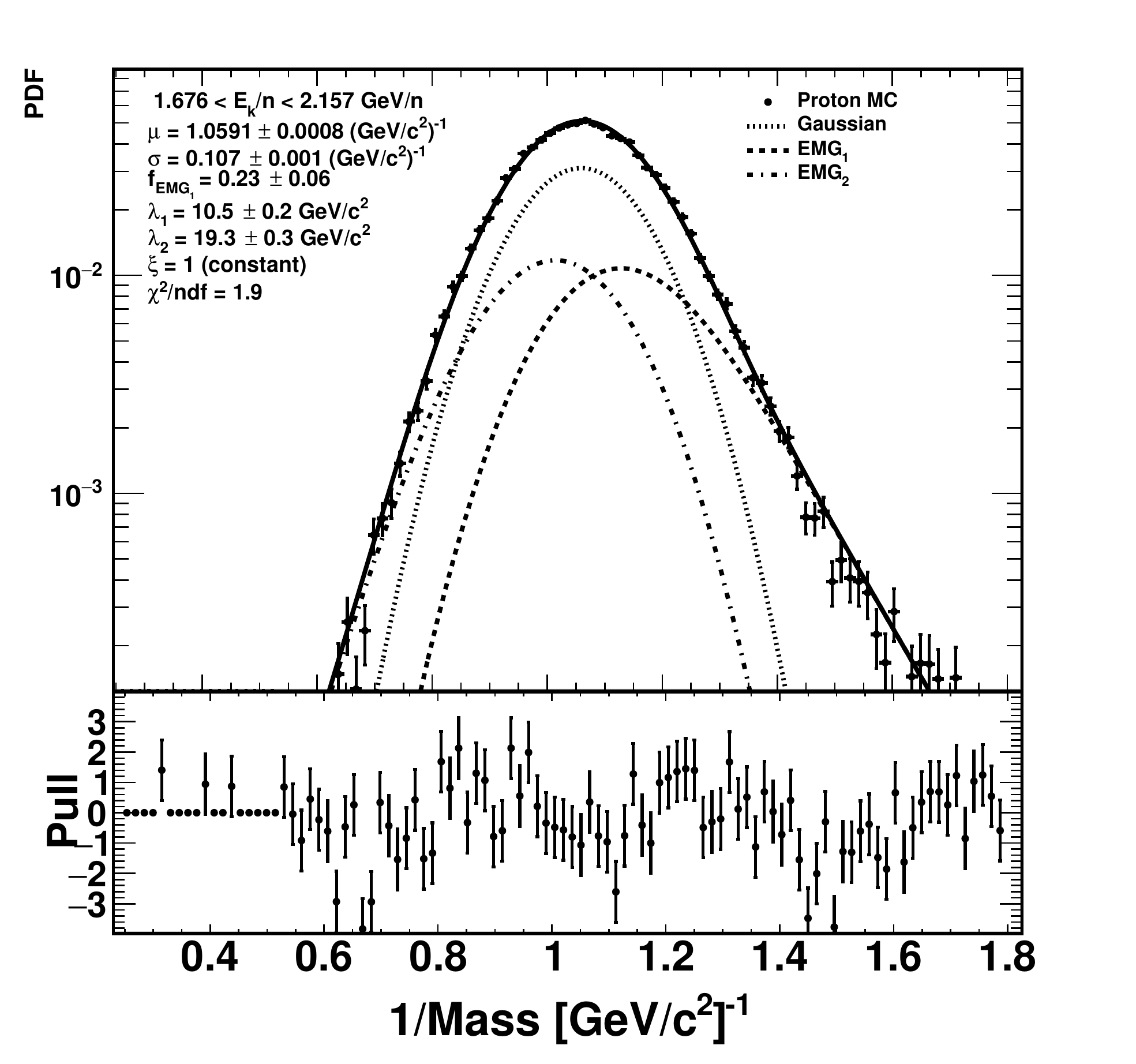}}
\caption{Example of fit of the parametric model to proton simulated events in the NaF range for events with kinetic energy per nucleon between 1.676 and 2.157 GeV/n (top panel). The different lines indicate the components of the model: the Gaussian is the thin-dashed line; the EMG for low mass tails is the thick-dashed; the EMG for high mass tails is the dotted-dashed line; the sum of all the contributions is the solid black. The bottom panel shows the pull of the fit.
  \label{fig:fit_mc_naf}
}
\end{figure}

\begin{figure}[!h]
\centering{\includegraphics[width=\linewidth]{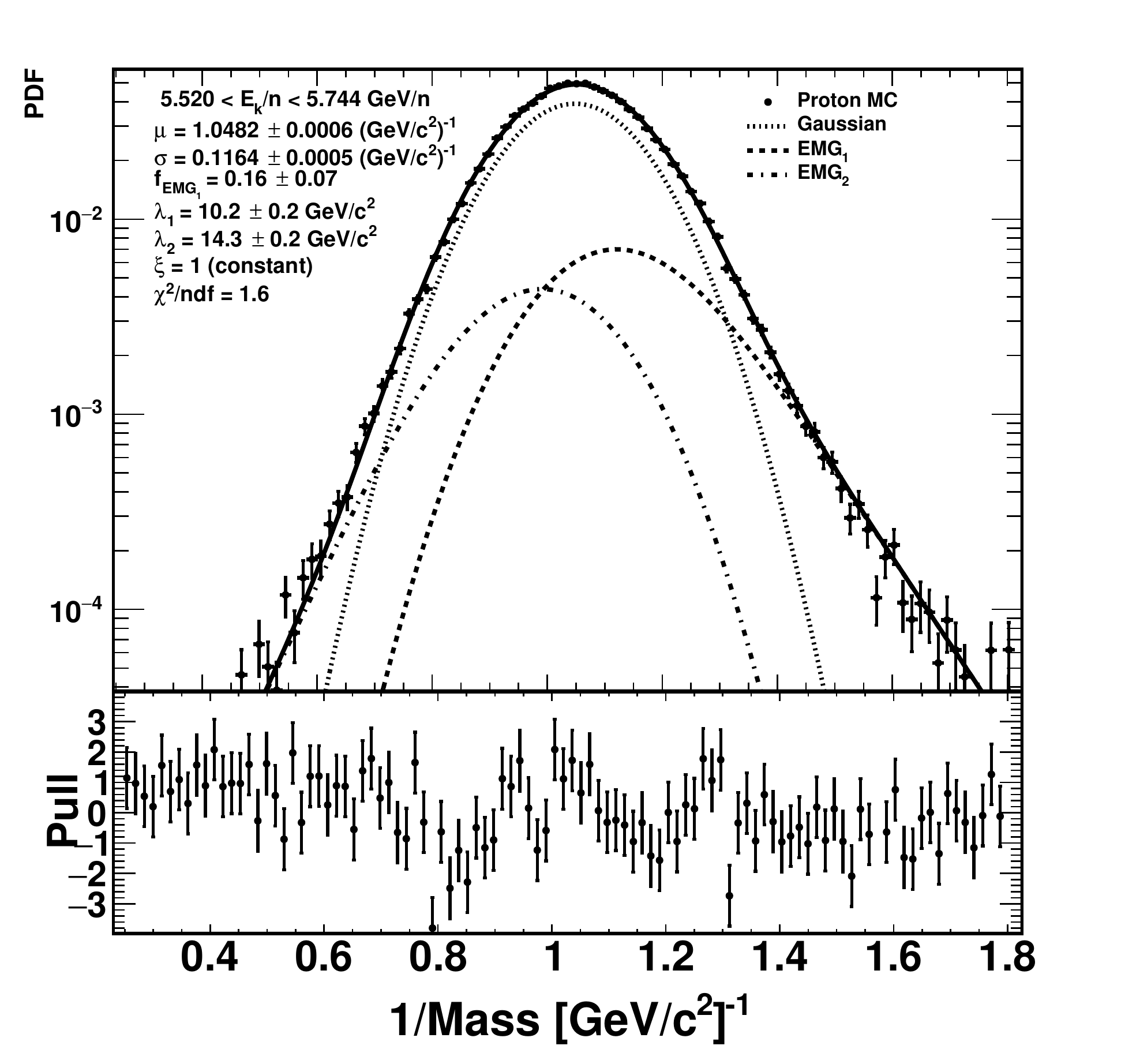}}
\caption{Example of fit of the parametric model to proton simulated events in the aerogel range for events with kinetic energy per nucleon between 5.520 and 5.744 GeV/n (top panel). The different lines indicate the components of the model: the Gaussian is the thin-dashed line; the EMG for low mass tails is the thick-dashed; the EMG for high mass tails is the dotted-dashed line; the sum of all the contributions is the solid black. The bottom panel shows the pull of the fit.
\label{fig:fit_mc_agl}
}
\end{figure}

Such interactions produce additional hits in the PMT plane of the RICH, causing the reconstructed ring to be larger, biasing the measured velocity and thus the mass towards higher values. Given this additional contribution, the parameterization of the RICH inverse mass requires the addition of another EMG, with an exponential tail to the left, to describe high-mass events, that is

\begin{multline}
M_{p}(1/m) = f_{G} \cdot G(1/m; \mu, \sigma) \\
 + f_{EMG_{1}} \cdot EMG(1/m; \mu, \xi \sigma, \lambda_{1})  \\
 + (1 - f_{G} - f_{EMG_{1}}) \cdot EMG(1/m; \mu, \xi \sigma, \lambda_{2})
\end{multline}

\noindent where $\mu$ is the mean, $\sigma$ the width if the Gaussian, $f_{EMG_{1}}$ is the fraction of the first EMG, $\lambda_{1}$ the low-mass exponential tail, $\lambda_{2}$ the high-mass exponential tail and $\xi$ the factor which determines the increase of the EMG width with respect to the width of the gaussian, as in the ToF. However, given the additional freedom given to the distribution by adding an extra EMG, $\xi$ has been set to 1 for simplicity. Examples of fits to proton MC in both the NaF and Aerogel velocity ranges are shown in figures \ref{fig:fit_mc_naf} and \ref{fig:fit_mc_agl}, respectively. Once again, it is clear that our parametrization can describe the proton MC distribution well, as stressed by the pull. Although they share a common model, given the differences of both radiator materials and their location inside AMS, the parameters of each radiator are different. Therefore they will be treated separately. Analogously to the case of the ToF, $f_{G}$ is also written as a function of other parameters and the expected inverse mass value. However, in this case, there are two EMGs with tails in opposite directions. Therefore the expected value is

\begin{multline}
    \langle1/m\rangle = f_{G}\cdot\mu + (f_{\mbox{EMG}_{1}})\cdot(\mu + 1/\lambda_{1}) \\
    (1-f_{G}-f_{\mbox{EMG}_{1}})\cdot(\mu - 1/\lambda_{2})
\end{multline}

\noindent which leads to the expression for $f_{G}$

\begin{multline}
f_{G} = 1 + \lambda_{2}\left[\langle1/m\rangle - f_{\mbox{EMG}_{1}}\cdot(\frac{1}{\lambda_{1}} + \frac{1}{\lambda_{2}}) - \mu\right]
\end{multline} 

\noindent where $\langle1/m\rangle$ is computed numerically from the inverse mass distribution.

\begin{figure}[!h]
  \centering{\includegraphics[width=\linewidth]{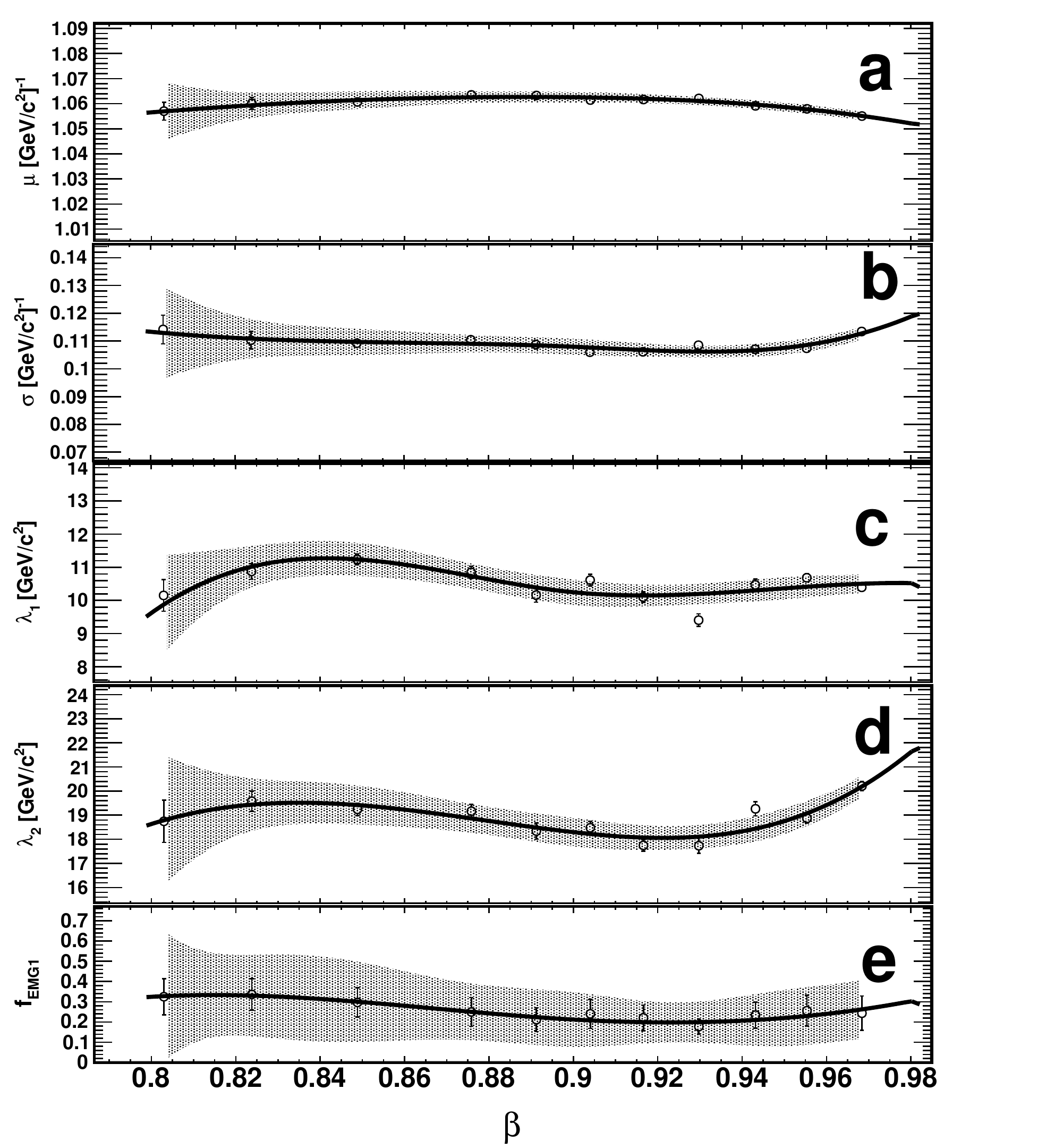}}
\caption[NaF parameters versus the velocity]{%
The parameters of the model as a function of $\beta$ in the NaF range. Panel (a) shows $\mu$; panel (b) shows $\sigma$; panel (c) shows $\lambda_{1}$; panel (d) shows $\lambda_{2}$, the parameter which describes the high-mass tails; and panel (e) shows $f_{EMG_{1}}$, the fraction of the low-mass tail EMG in the model. In all panels, the black curve represents the spline fit and the gray band the 95$\%$ confidence interval of the fit.
  \label{fig:pars_naf}

}

\end{figure}

A typical example of fit to proton MC is shown in figure \ref{fig:fit_mc_naf}, together with the pull. The necessity of a third component to account for high-mass tails is evident by looking at the region with $1/m < 0.8 \, (\text{GeV}/c^{2})^{-1}$, where it is possible to see that the proton MC is shallower than a Gaussian distribution. Figure \ref{fig:pars_naf} shows the evolution of the parameters with the velocity. Panel (a) shows $\mu$, which is close to what is expected for protons, $1.06 \, (\text{GeV}/c^{2})^{-1}$, in most of the range, with the deviation at lower velocities coming from the energy loss particles undergo while crossing the detector; panel (b) shows $\sigma$, which displays a constant behavior at the beginning of the range, increasing with the velocity, above $\beta > 0.95$, as a result of the $\gamma^{2}$ dependence of the mass resolution; panel (c)  shows $\lambda_{1}$, which has a constant behavior with velocity apart from the beginning of the range, where the Cherenkov emission threshold affects the distribution; panel (d) shows $\lambda_{2}$, the parameter responsible for the high-mass tail. Similarly to $\lambda_{1}$, it has a drop at the beginning of the range due to the Cherenkov-emission threshold. Panel (e) shows $f_{EMG_{1}}$, which also has a constant behavior with the velocity, indicating that the high-inverse mass (low mass) tails are independent of the velocity.

The parametrization is the same for both the NaF and the AGL. Still, the radiators differ in several aspects: the AGL reaches higher velocities, allows for a better velocity resolution, and due to the different geometry, has an acceptance that is tenfold that of the NaF. Hence, the values of the parameters are different.

An example of the fit to the proton simulated events in the AGL radiator is shown in figure \ref{fig:fit_mc_agl}. The top panel shows the inverse mass distribution with the different components, while the bottom panel shows the pull, which indicates an excellent agreement between the proposed model and simulations. As in the NaF range, the importance of adding a component to account for the high-mass tail is also clear from the increase of the width of the
distribution below $1/m  = 0.8$ $(\text{GeV}/c^{2})^{-1}$.

\begin{figure}[!h]
 \centering{\includegraphics[width=\linewidth]{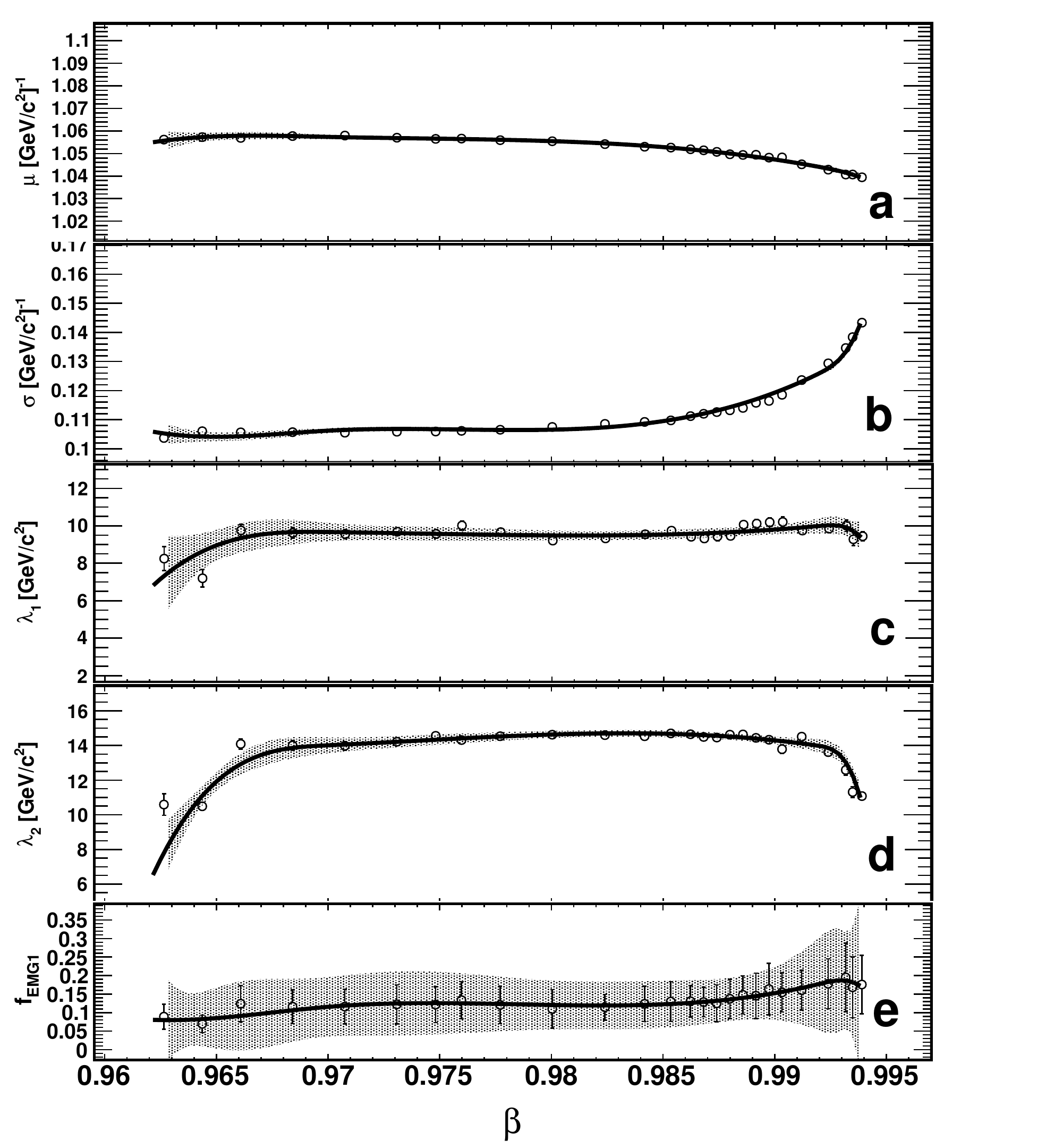}}
\caption[AGL parameters versus the velocity]{%
The parameters of the model as a function of $\beta$ in the AGL range. Panel (a) shows $\mu$; (b) shows $\sigma$;  (c) shows $\lambda_{1}$; (d) shows $\lambda_{2}$, the exponential tail of the high-mass EMG; (e) shows $f_{EMG_{1}}$; In all panels, the black curve represents the spline fit and the gray band the 95$\%$ confidence interval of the fit.
\label{fig:pars_agl}
}

\end{figure}
Figure \ref{fig:pars_agl} shows the evolution of the parameters with the velocity for the AGL. Similarly to the NaF, $\mu$, shown in panel (a), is compatible with what is expected for the proton mass; panel (b) shows the width, $\sigma$, which behaves the same way it does in the NaF, increasing rapidly with the velocity; panel (c) shows the low-mass tail, described by $\lambda_{1}$, which displays a flat behavior within the range (except near the Cherenkov threshold), which is consistent with the flat behavior of the rigidity resolution in this region;  panel (d) shows $\lambda_{2}$, which, on the other hand, has a stronger velocity dependence. It starts with lower values (meaning a more prominent tail) at low velocities due to the existence of the Cherenkov emission threshold. At velocities close to the threshold ($\beta_{Th} \approx 0.96$), the signal is feeble and thus the reconstruction is more sensitive to the background. Well beyond the threshold, $\lambda_{2}$ increases (less prominent tails) to a stable value when it has a sharp drop, meaning the high-mass tails are more present at $\beta > 0.99$. This comes effect comes from the increase of the diameter of the Cherenkov rings with the velocity, which increases the probability of losing a fraction of the signal in the ECAL hole; panel (e) shows $f_{EMG_{1}}$, which decreases with the velocity. This behavior is explained by the fact that the high-mass tails become more critical at higher velocities hence one of the EMGs loses weight, while the other increases.

\subsection{Deuteron templates}

Following the construction of the proton template, the deuteron templates were built through a scaling factor applied to the proton parameters and the addition of a fragmentation component.

\subsubsection{Scaling of the parameters}

The scaling of the parameters is related to the difference between the mass of the isotopes. Since $\mu$ represents the expected value of the inverse mass, $\mu_{d}/\mu_{p} \sim m_{p}/m_{d}$, hence $\mu_{d} = \mu_{p} \cdot \alpha_{d}$, where $\alpha_{d} = m_{p}/m_{d} \approx 0.5$. The mass resolution of protons and deuterons is approximately the same because the rigidity resolution is essentially flat below 20 GV. In addition, the resolution of $1/m$ and $m$ are the same, that is $\sigma_{m}/m = \sigma_{1/m}/(1/m)$. Therefore we can write $\sigma_{p}/(1/m_{p}) \sim \sigma_{d}/(1/m_{d}) $, hence $\sigma_{d} = \alpha_{d} \sigma_{p}$. A similar reasoning can be applied to the tail parameters $\lambda_{1}$ and $\lambda_{2}$. The variance of an EMG, Var, is given by two components, one from the Gaussian and another from the exponential, $\mbox{Var} = \sigma^{2} + 1/\lambda^{2}$.  For deuterons, the variance is written can be written as a function of the parameters from the proton template as $\mbox{Var}_{d} = \alpha_{d}^{2} \sigma_{p}^{2} + (\kappa/\lambda_{p})^{2}$, where $\kappa$ is the scaling factor to be determined. Using the same reasoning applied to $\sigma$, we have that $\mbox{Var}_{d} \, m_{d}^{2} \sim \mbox{Var}_{p} \, m_{p}^{2}$ , which leads to $\kappa = 1/\alpha_{d}$. Hence, we have that $\lambda_{d} = \lambda_{p}/\alpha_{d}$.  Table \ref{table:scaling} summarizes how the parameters are scaled for the deuteron model.

\begin{table}[]
\centering{%
\begin{tabular}{ccc}
\hline
p             &               & d                        \\ \hline
$\mu$         & $\rightarrow$ & $\alpha_{d} \mu$         \\
$\sigma$      & $\rightarrow$ & $\alpha_{d} \sigma$      \\
$\lambda_{1}$ & $\rightarrow$ & $\lambda_{1}/\alpha_{d}$ \\
$\lambda_{2}$ & $\rightarrow$ & $\lambda_{2}/\alpha_{d}$ \\ \hline
\end{tabular}
\caption{Relationship between the parameters of the proton model (left column) and the deuteron model (right column) through the scaling factor.}
\label{table:scaling}
}
\end{table}
 
 The other parameters present in the model, such as $f_{G}$ and $f_{EMG_{1}}$, are kept to the same values obtained for the protons, without scaling, because they do not directly alter the shape of the distribution but rather dictate the fraction of each component. Considering that the other parameters are scaled by a common factor, it is reasonable to expect that the relative weights of each distribution remain the same. Finally, $\xi$ is also not scaled because it simply dictates how much larger is the width of the EMG as compared to the gaussian core, which is already scaled by $\alpha_{d}$.

\subsubsection{Deuteron-to-proton fragmentation}

\begin{figure}[h]
\centering{\includegraphics[width=\linewidth]{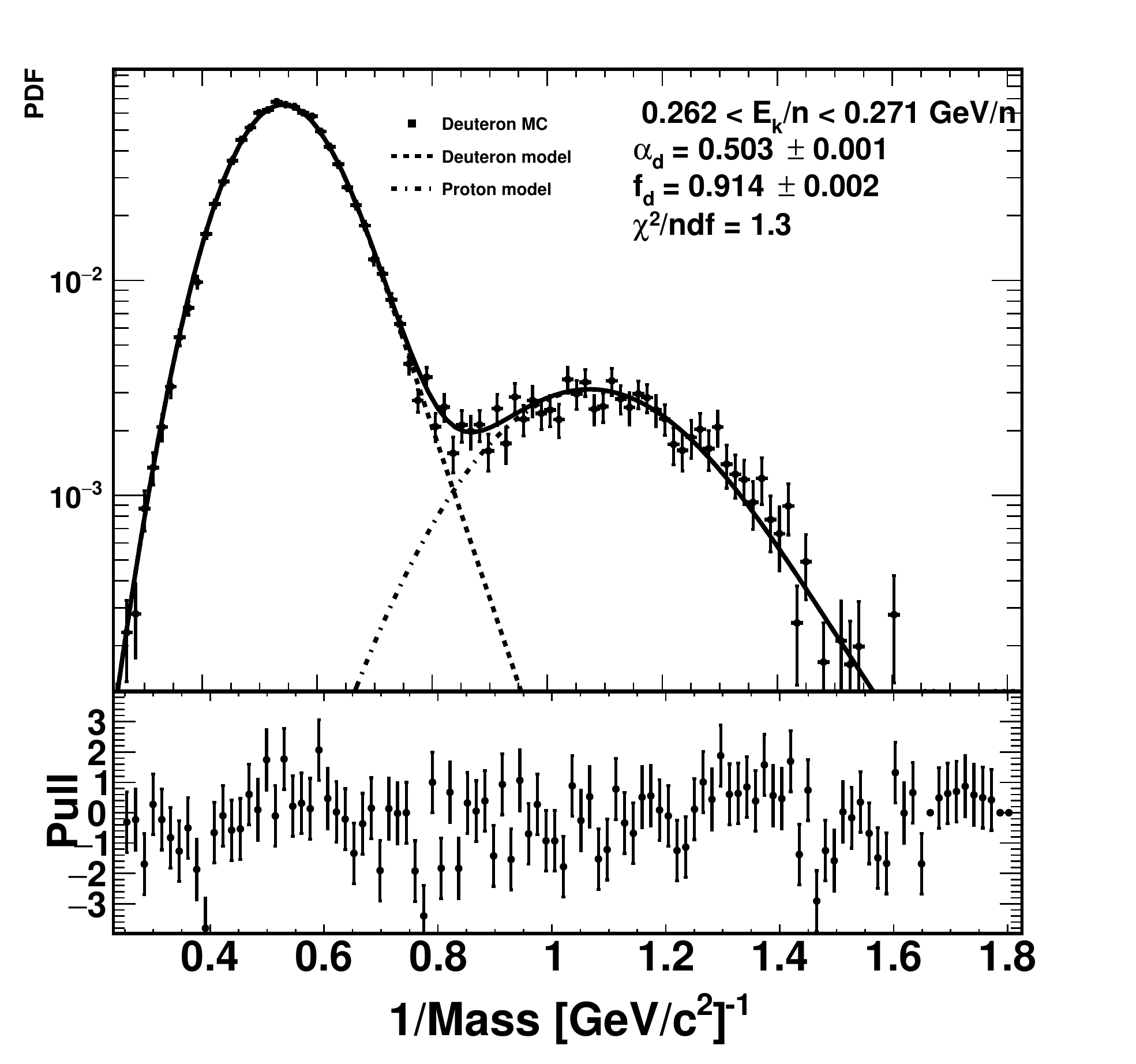}}
\caption{\label{fig:fit_d_tof} Example of fit to deuteron simulated events in the ToF range, for events with kinetic energy per nucleon between 0.262 and 0.271 GeV/n. The dashed curve represents the deuteron component; the dot-dashed curve represents protons coming from d $\rightarrow$ p fragmentation, and the solid line is the sum of both. The bottom panel shows the pull of the fit.}
\end{figure}

During their interactions with AMS material, deuterons can fragment and produce protons. This effect is clearly shown in figure \ref{fig:fit_d_tof}, where the inverse mass distribution of simulated deuteron events in the ToF range is shown: besides the main deuteron peak, with an inverse mass around $0.5 \, (\text{GeV}/c^{2})^{-1}$, there is a fraction of events around $1 \, (\text{GeV}/c^{2})^{-1}$.

Hence, to create a complete model for deuterons ,$M_{d}^{\text{\tiny Total}}$, a proton component is added to the deuteron PDF, $M_{d}$, thus giving:

\begin{equation}
    M_{d}^{\text{\tiny Total}}(1/m) = f_{d} \, M_{d}(1/m) + (1 - f_{d} )  M_{p}(1/m)
    \label{eq:dfrag}
\end{equation}

\noindent where $f_{d}$ represents the fraction of deuterons that did not fragment, and $M_{p}$ is the proton model. Hence, in order to obtain a complete deuteron model, which also takes into account the fragmentation, only $\alpha_{d}$ and $f_{d}$ must be obtained. To this end, the procedure is as follows: the proton template parameters are fixed to the values obtained from the fit to proton simulated events, and the deuteron template is built by scaling the parameters according to table \ref{table:scaling}. Both templates are then summed according to equation \ref{eq:dfrag}, resulting in the total model, which is then fitted to deuteron simulated events to extract the values of both $\alpha_{d}$ and $f_{d}$.

This is done with the same velocity bins as the one used for protons, such that the velocity dependence of these parameters can be studied.

An example of the fit to simulated deuteron events is shown in figure \ref{fig:fit_d_tof}. It is clear the scaling works remarkably well; moreover, the fit result for the scaling parameter, $\alpha_{d}$, is around 0.5 as expected. The fragmentation component is also evident: in this particular bin, with kinetic energy per nucleon between 0.262 and 0.271 GeV/n, around $8\%$ of the deuterons are fragmented and were identified as protons. 

\begin{figure}[!h]
\centering{\includegraphics[width=\linewidth]{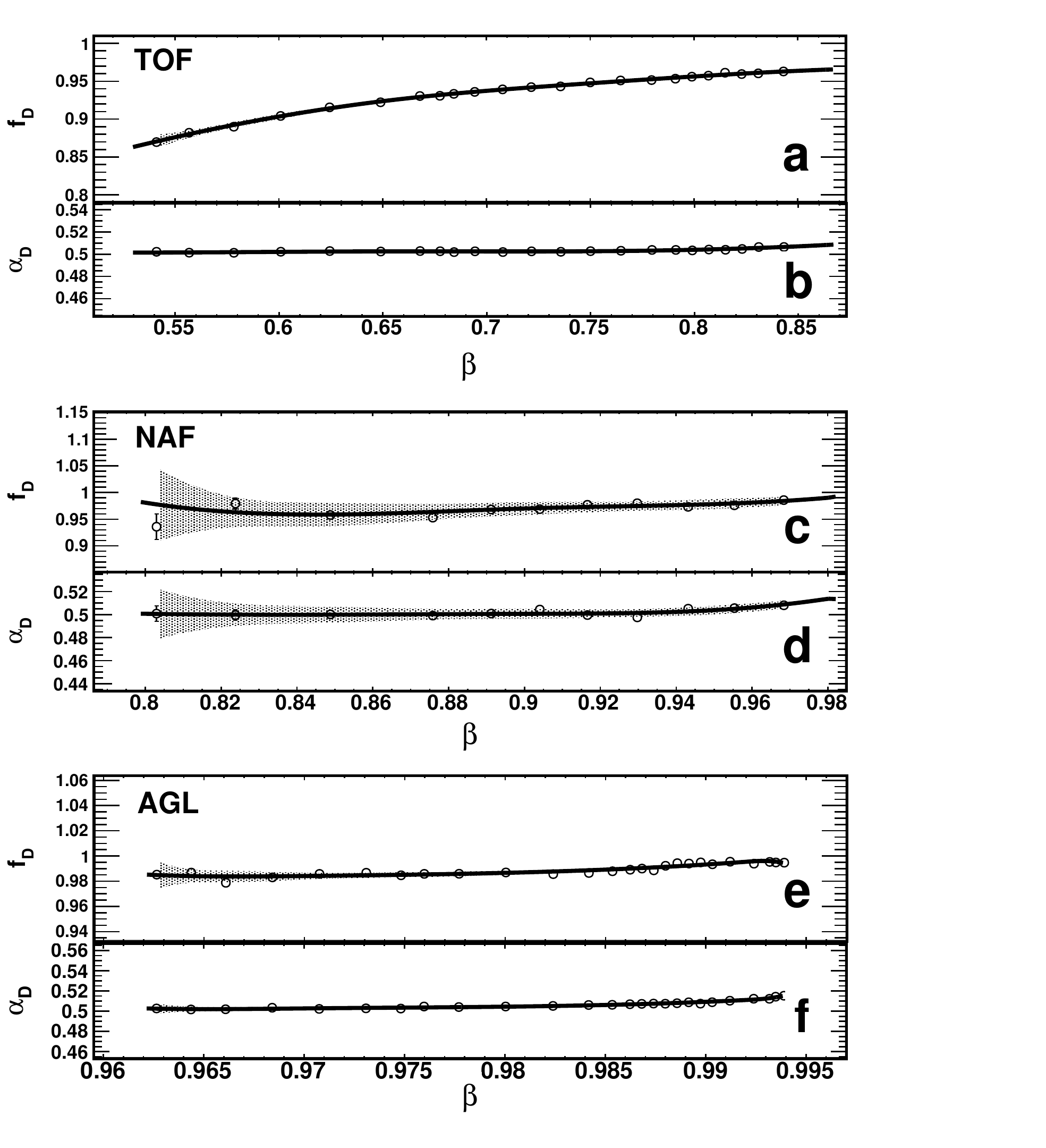}}
\caption{Evolution of deuteron model parameters with the velocity. Panels (a) and (b) show the scaling factor, $\alpha_{d}$ and the fraction of non-fragmented deuterons, $f_{d}$, for the ToF. Similarly, the same parameters are shown for the NaF range in panels (c) and (d) and AGL in panels (e) and (f). The solid-black lines correspond to the spline fits, while the gray bands correspond to the $95\%$ confidence interval of each fit.}
\label{fig:deuteron_pars}
\end{figure}

Figure \ref{fig:deuteron_pars} shows the velocity dependence of the parameters of the deuteron model. Panels (a), (c), and (e) display $f_{d}$, the fraction of non-fragmented deuterons, in the ToF, NAF, and AGL range, respectively. It reaches its lowest value, about 0.85, at the lowest velocities in the ToF. It steadily increases with velocity until the beginning of the NaF range, remaining constant after that. Panels (b), (d), and (f) display $\alpha_{d}$ for the three ranges; as expected, the value remains stable around 0.5 in all three velocity intervals. 

\subsection{Helium fragmentation background}

\begin{figure}[h]
\centering
\includegraphics[width=\linewidth]{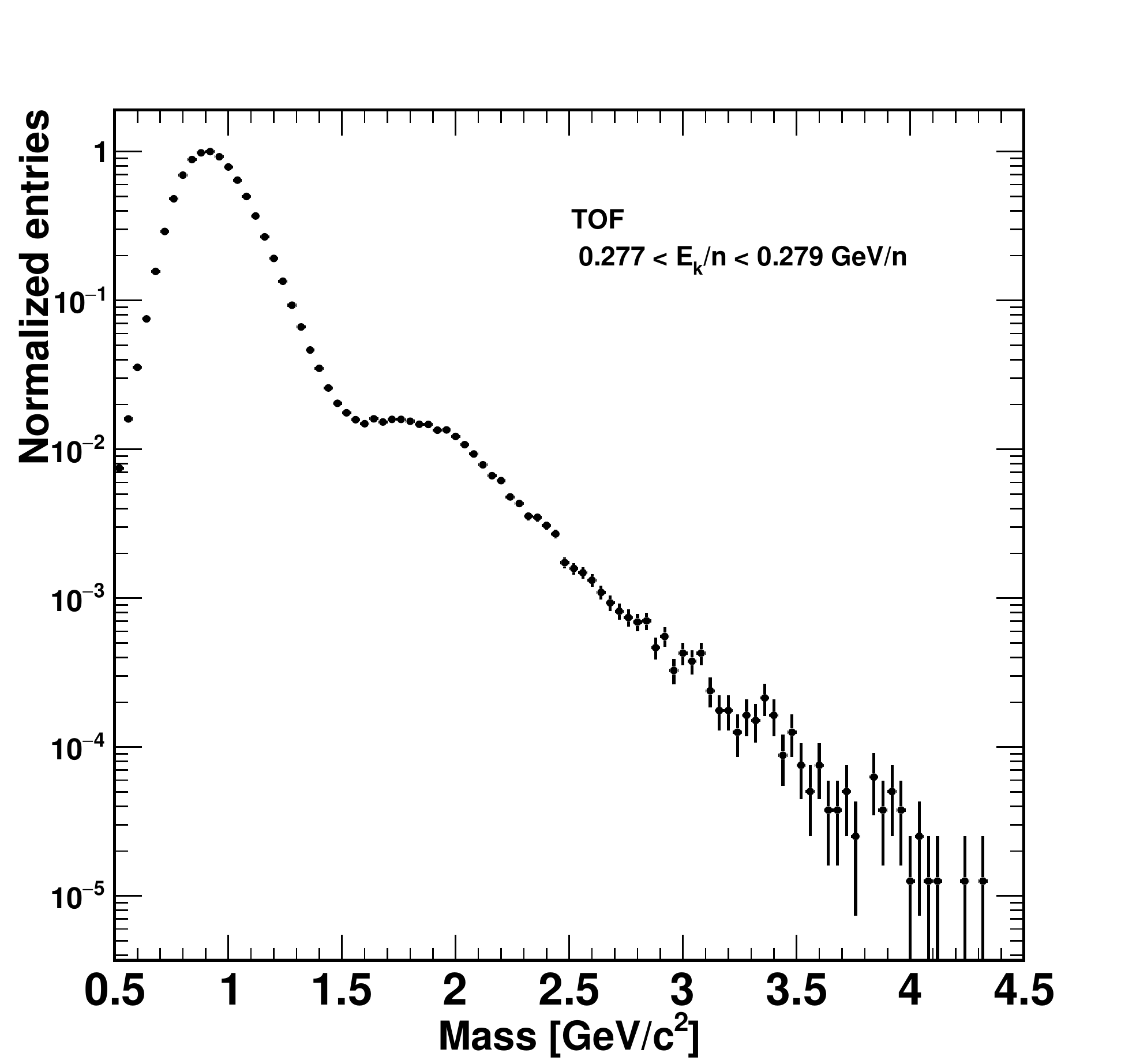}
\caption{\label{fig:mass_triton} Mass distribution of experimental data in the ToF range for events with kinetic energy per nucleon between 0.277 and 0.279 GeV/n. The proton peak and deuteron shoulder are clearly seen around 1 and 2 GeV/n, respectively. Tritons coming from $^{4}$He fragmentation populate the high-mass region.}
\end{figure}

Helium amounts to about $10\%$ of CRs. Given this abundance, the background of such particles fragmenting while passing through AMS material, leaving spurious $Z = 1$ signals, cannot be neglected. It is known that $^{4}\text{He}$ is responsible for about $80\%$ of the helium flux in CRs \citep{ams-helium-isotopes}. Hence, this species is expected to be the largest contributor to this type of background. The fragmentation of such particles while interacting with $^{12}$C, the main material of the AMS-02 detector \citep{qiyan}, has been studied by several nuclear physics experiments, and they show that not only protons and deuterons are produced, but also tritons \citep{HeFrag}\citep{HeFragNoguchi}. Although this component is small (below $1\%$), it can still be seen in experimental data as a high-mass tail to the right of the deuteron shoulder in the mass distribution, as shown in figure \ref{fig:mass_triton}.

The fiducial volume of the mass measurement, that is, the region of the detector where it is performed, is the one from the upper ToF to the RICH detection plane. In this volume, the velocity and rigidity measurements are performed; hence, any fragmentation occurring above the upper ToF can contaminate the sample. Aiming to mitigate this effect, a cut selecting single-charged particles is applied to the first layer of the tracker (L1). However, the sample can still have some background due to the following mechanisms:

\begin{enumerate}
    \item \textbf{Fragmentation before the first layer of the tracker:} it is possible that some particles fragment by interacting with material above the first layer, such as the low-energy radiation shield and the support of L1.
    \item \textbf{Incorrect charge measurement at the first layer of the tracker (L1):} given that the single-layer charge resolution of the AMS tracker is about $8\%$ for single-charged particles \citep{yijia_charge}, it is possible that some $Z = 2$ events survive the selection, fragmenting further in the detector, such as the TRD or the support of the upper ToF.
\end{enumerate}

Tightening the cut on the L1 charge would reduce the background coming from point 2 but would reduce the data sample size considerably and would not remedy the background coming from the first point. Therefore, such contamination is unavoidable, and it must be estimated at the level of the signal extraction, as described in the next section.

\subsection{Helium fragmentation templates}

Given the species produced by the fragmentation of helium, the construction of the fragmentation template starts by obtaining the PDF of tritons. Similar to the case of deuterons, a triton template can be obtained by scaling the parameters of the proton template with a factor $\alpha_{t}$, which in this case is approximately 1/3.

With the triton template, the fragmentation model can be written as:

\begin{multline}
M_{\text{He} \rightarrow X}(1/m) = f_{\text{He}\rightarrow p} \, M _{p}(1/m) +   f_{\text{He}\rightarrow d} \, M_{d}(1/m) \\ 
+ (1 - f_{\text{He}\rightarrow p} - f_{\text{He}\rightarrow d})  M_{t}(1/m)
\label{eq:frag_model}
\end{multline}

where $f_{\text{He}\rightarrow p}$ and $f_{\text{He}\rightarrow d}$ represent the fractions of He which fragmented into protons and deuterons, respectively.

Hence, the estimation of each of these fractions, as well as of $\alpha_{t}$, is all that is left for the construction of the full fragmentation model. This could be done, in principle, by using He simulated events that passed a selection that requested the electric charge to be compatible with 1. However, this can also be done in a data-driven way, which has the advantage of not relying on any sort of cross-section modelling.

\begin{figure}[h]
\centering
\includegraphics[width=\linewidth]{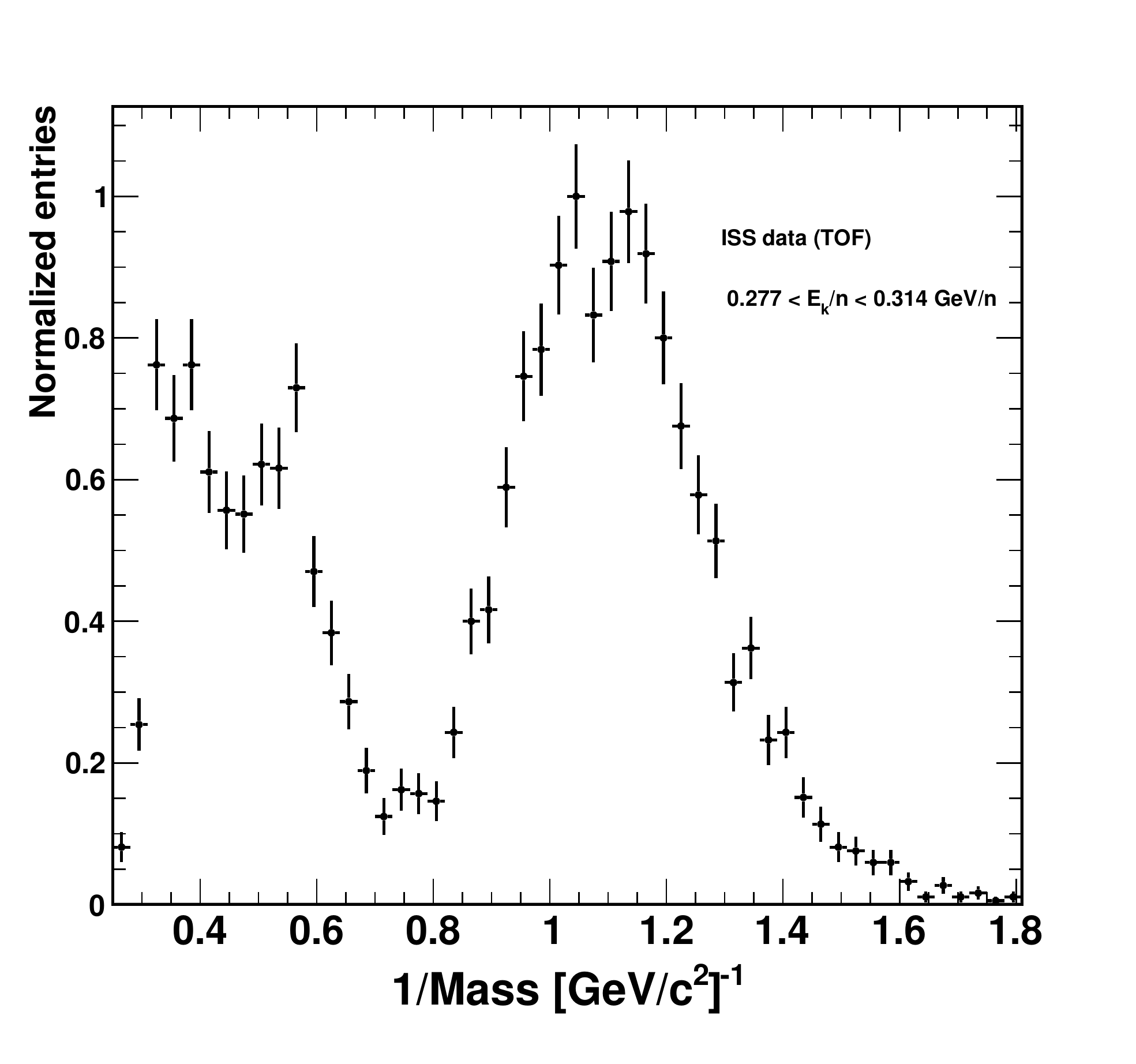}
\caption{\label{fig:frag_example} Example of the fragmentation sample in the ToF range, for events with kinetic energy per nucleon between 0.277 and 0.314 GeV/n.}
\end{figure}

For this method, we assume $f_{He\rightarrow p}$ and $f_{He\rightarrow d}$ are independent of where the fragmentation took place, as long as it has occurred before the fiducial volume of the analysis; therefore we must ensure the incoming particle in an helium nucleus that fragmented inside AMS, by requiring $Z = 2$ above the upper ToF and $Z=1$ below it. This is done by requiring $Z$ to be in the range $1.6 < Z < 2.4$ at L1 and in the TRD; and $0.6 < Z < 1.4$, in the inner tracker. An example of the inverse mass distribution after this selection is shown in figure \ref{fig:frag_example}, where the peaks of the three isotopes can be seen. With this sample, the fragmentation fractions can be obtained as a function of velocity for each range. However, it is important to stress that such selection is very restrictive; therefore, the number of events per analysis bin is not always enough to extract the parameters with uncertainty at the percent level, especially in the NaF range, where the geometric acceptance is significantly smaller. Therefore, the bins were grouped to increase the statistics available for each fit to remedy this effect.

\begin{figure}[!h]
\centering
\includegraphics[width=\linewidth]{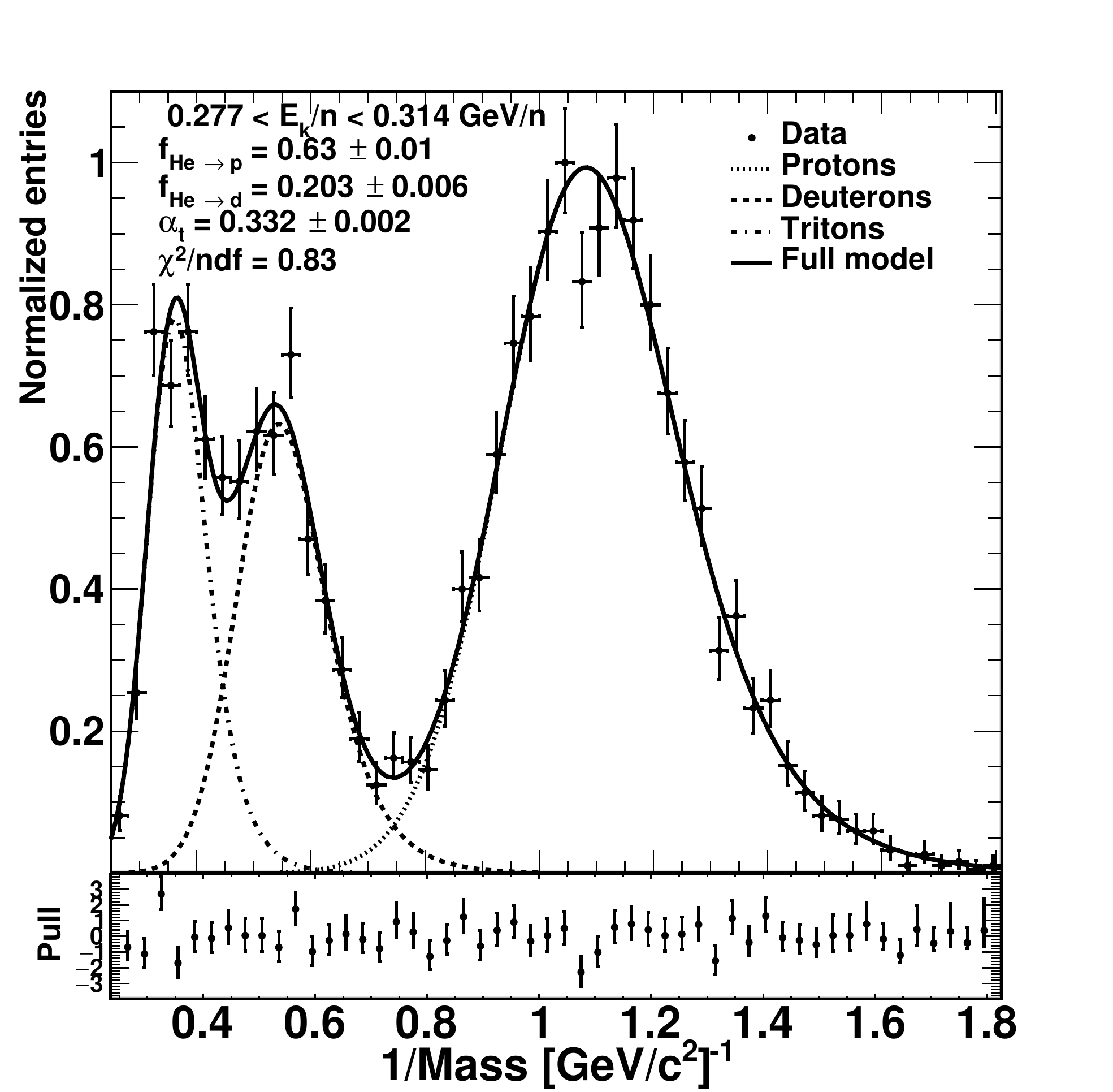}
\caption{\label{fig:fit_frag} Example of fit of the fragmentation model to experimental data in the ToF range, for events with kinetic energy per nucleon between 0.277 and 0.314 GeV/n. The different lines indicate the components of the model: the protons are the thin-dashed line; the deuterons are the thick-dashed; the tritons are the dotted-dashed line; the sum of all the contributions is the solid black. The bottom panel shows the pull of the fit.}
\end{figure}

Similar to what has been done in the case of the deuteron template, the parameters are obtained by performing a fit where the ones of the other species (deuterons and protons) are fixed according to what has been obtained in the previous fits to these species. All that is left to be adjusted are $\alpha_{t}$ and the fragmentation fractions. Figure \ref{fig:fit_frag} shows an example of fit of the model described in equation \ref{eq:frag_model} to experimental data. As indicated by the pull, the agreement between our model and the data is excellent.

\begin{figure}[!h]
\includegraphics[width=1\linewidth]{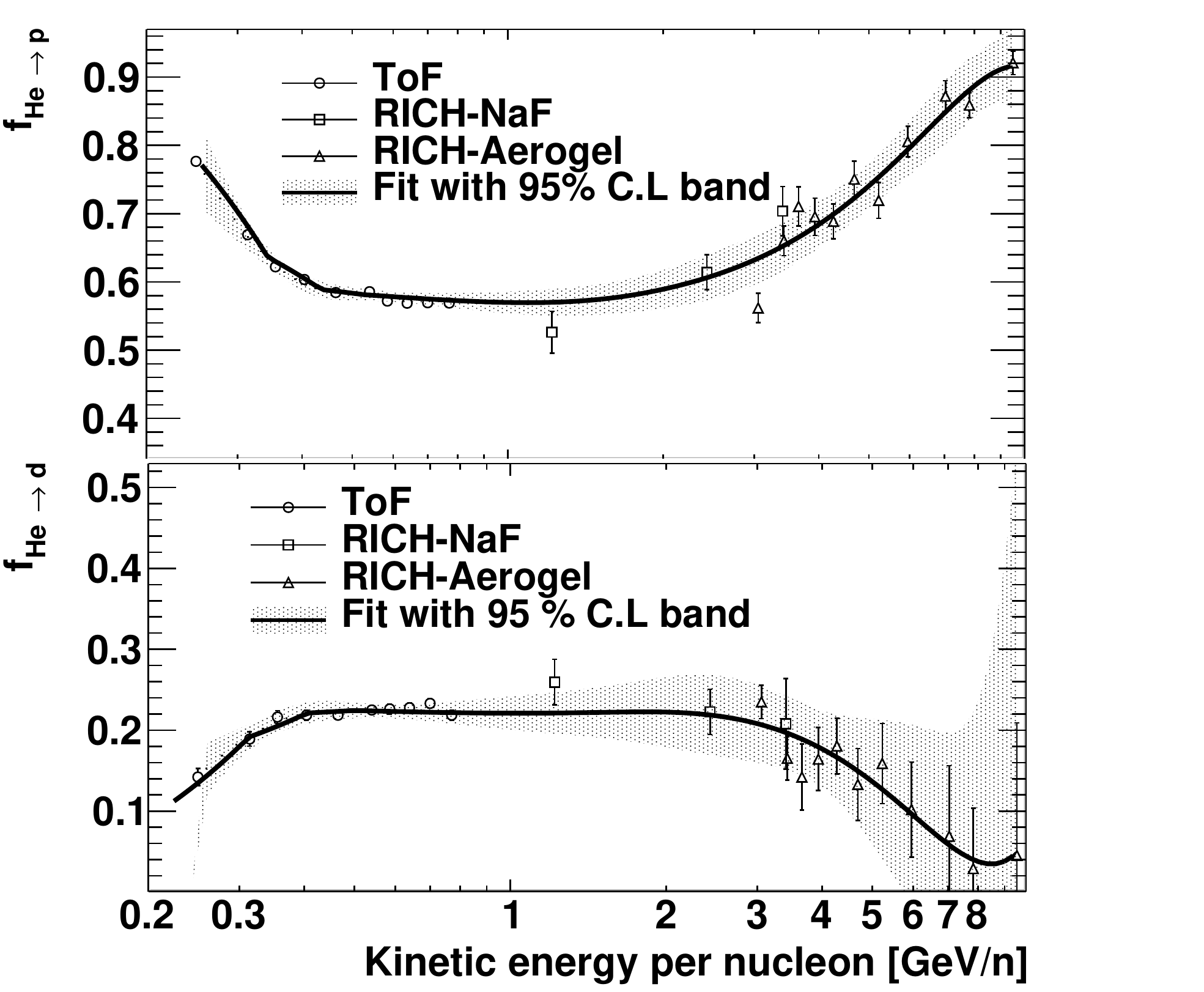}
\caption[]{\label{fig:hefrag_pars} The top and bottom panels show $f_{He\rightarrow p}$ and $f_{He\rightarrow d}$ versus the kinetic energy per nucleon, respectively. The three different markers correspond to the different subdetectors; the black curve represents the spline fit and the gray band the 95$\%$ confidence interval of the fit.}
\end{figure}

Since the bins are grouped to have enough events for the extraction of the parameters, there are fewer points per subdetector range; therefore, building curves that describe the velocity dependence was not be done separately for each subdetector as it was in the previous cases. However, since the goal is to extract the fractions $f_{He\rightarrow p}$ and $f_{He\rightarrow d}$, the three measurements can be combined: the physical process which dictates the values of these fractions is the fragmentation of $^{4}$He, and there are no reasons to believe it is not continuous with the velocity. Hence, in the velocity ranges where there is overlap between the detectors,  both measurements of $f_{He\rightarrow p}$ and $f_{He\rightarrow d}$ must be compatible. Consequently, a spline is fitted to the three ranges simultaneously, building a global model for the fractions versus the velocity. Figure \ref{fig:hefrag_pars} shows $f_{He\rightarrow p}$ and $f_{He\rightarrow d}$ as a function of the kinetic energy.

It is possible to note that the values of the fractions measured in the different velocity ranges are compatible in both cases. The ToF and Aerogel are less penalized by the loss of statistics, while the NaF provides only 3 points, reinforcing the necessity of building a global model.

The parameter $\alpha_{t}$ was found to be constant with velocity, with value $\alpha_{t} = 0.332 \pm 0.002$, as shown in figure \ref{fig:fit_frag}. Having obtained all the necessary parameters for the template model, the identification of the isotopes in data was performed as it will be discussed in the next section.

\section{Fit to experimental data}

The experimental data sample consists of several components: unfragmented protons and deuterons; protons from deuteron fragmentation; and protons, deuterons, and tritons from helium fragmentation. The complete model, $M_{\text{Total}}$, is then given by

\begin{multline}
M^{\text{Total}}(1/m) = f_{1} M_{p} + f_{2} M_{d} + f_{3} M_{d \rightarrow p} + \\
f_{4} M_{\text{He}\rightarrow p} + f_{5} M_{\text{He}\rightarrow d} + f_{6} M_{\text{He}\rightarrow t}
\end{multline}

\noindent where $\sum_{n=1}^{n=6} f_{n} = 1$. Using the models built before for the deuterons and fragmented helium, the notation can be simplified as follows

\begin{multline}
M^{\text{Total}}(1/m) = f_{1} M_{p} + \underbrace{f_{2} M_{d} + f_{3} M_{d \rightarrow p}}_{M_{d}^{\text{Total}}} + \\
\underbrace{f_{4} M_{\text{He}\rightarrow p} + f_{5} M_{\text{He}\rightarrow d} + f_{6} M_{\text{He}\rightarrow t}}_{M_{\text{He}\rightarrow X}}
\end{multline}

\noindent therefore, the total model is given by

\begin{multline}
M^{\text{Total}}(1/m) = f_{p} M_{p} + f_{d} M_{d}^{\text{Total}} + f_{\text{He}} M_{\text{He}\rightarrow X}
\label{eq:total_model}
\end{multline}

\noindent where $f_p + f_d + f_{He} = 1$.

The fit to the data is performed using bins of velocity, thinner than those used for the simulated data set due to the abundant number of events. To have the correct parameter values for the model in every bin, the splines describing each of the model parameters were used to obtain the templates. As with any other fitting technique, there are uncertainties that must be taken into account. To that end, the uncertainty associated with each parameter of the model is given by the gray band around the fits, which represents the $95 \%$ confidence interval. These uncertainties on the parameters are then propagated to the fits to data by using nuisance parameters. In every bin, a Gaussian constraint is created for each parameter. The mean value is the evaluation of the spline at the bin center and the width corresponding to the uncertainty on that parameter calculated in that given bin. These constraints are then added to the likelihood to be maximized during the fit procedure. In effect, that means the parameters can change during the fit, but the Gaussian constraints limit their freedom. Since the parameters can fluctuate during the fit, there is no need to tune the simulations by smearing and shifting the mass distribution to find the best fit. It also means that the final uncertainty of the fit already incorporates these variations.

\begin{figure*}[!h]
\includegraphics[width=\linewidth]{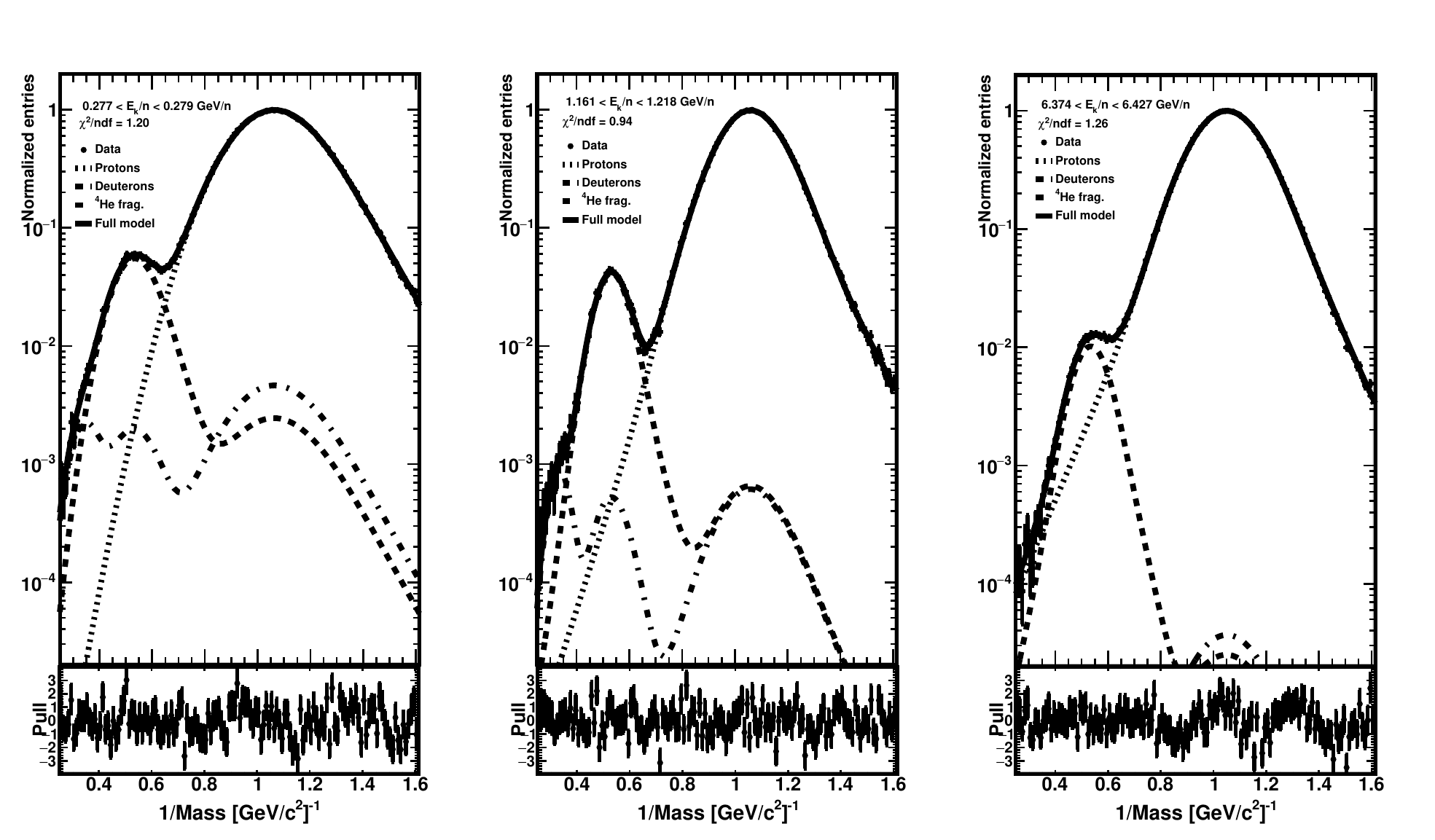}
\caption{Fit of the parametric model experimental data in the ToF (left), RICH-NaF (middle), and RICH-Aerogel (right) ranges. The different lines indicate the components of the model: the protons are the thin-dashed line; the deuterons are the thick-dashed; the fragmented are dotted-dashed line; the sum of all the contributions is the solid black. The bottom panels show the pull of each fit.}
\label{fig:fits_data}

\end{figure*}

Figure \ref{fig:fits_data} shows examples of fits to experimental data in the three velocity ranges, displaying the behavior of the fit in low, intermediate, and high energies. In the three cases, the pull shows an excellent agreement between the data and the model. It is also important to note that including the fragmentation components, especially the one from helium, is essential for the proper description of the data . Without it, the model would not be able to reproduce the data for inverse-masses below 0.4 $(\text{GeV}/c^{2})^{-1}$.

\subsection{Mass resolution}

\begin{figure}[!h]
\centering
\includegraphics[width=\linewidth]{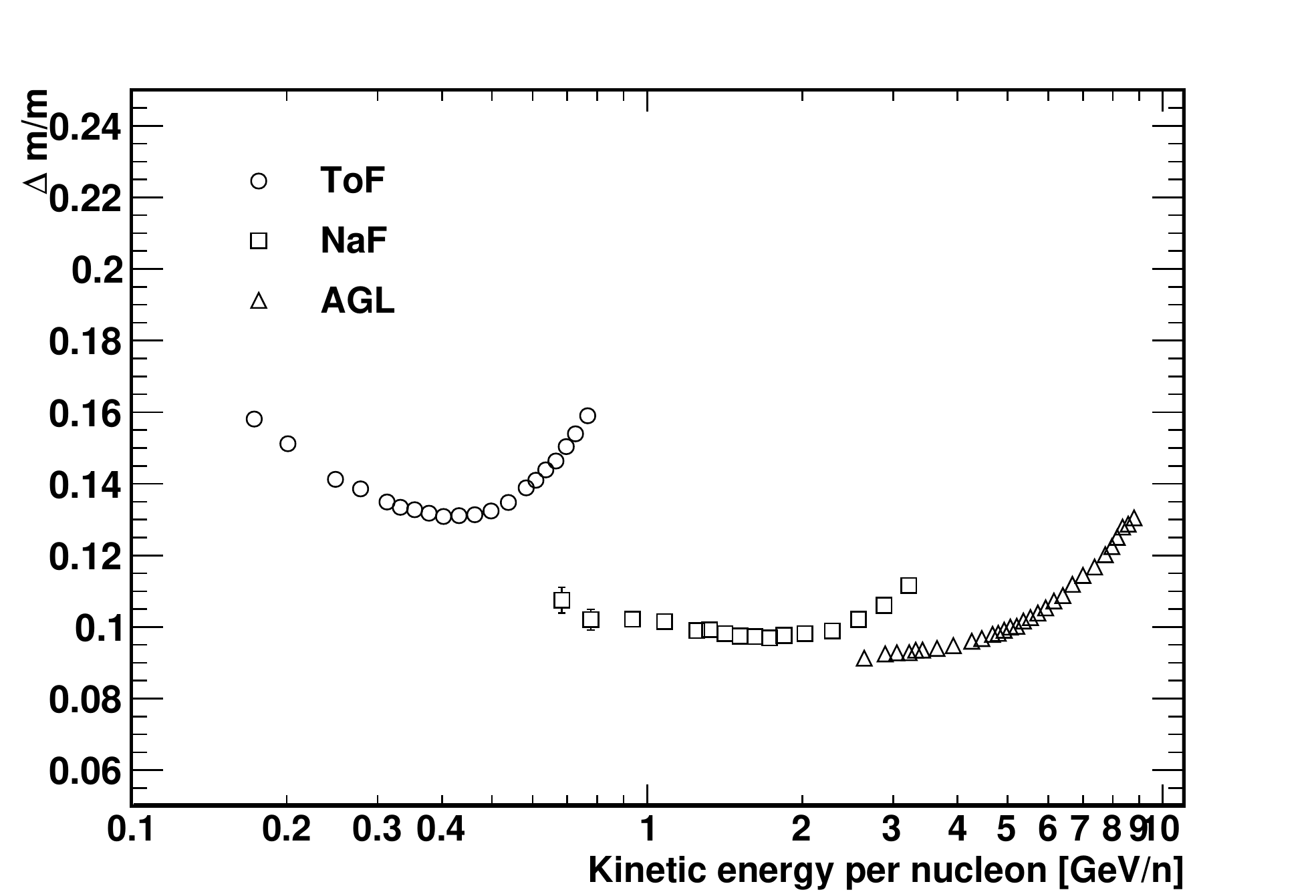}
\caption{\label{fig:massres} Mass resolution, $\Delta m/m$ as a function of the kinetic energy per nucleon as extracted from experimental data. Circles represent the ToF; squares the NaF; triangles the AGL. See text for discussion.}
\end{figure}

As the parameters were allowed to fluctuate within the uncertainties of the spline fits, besides the counts of each particle type, the fits to experimental data yield new values of the parameters of the model itself. In particular, the model was built in such a way that $\mu / \sigma$ is a good estimation of the detector mass resolution, $\Delta m / m$, due to the fact that the Gaussian is the largest component of the model, as seen in the fits shown in figures \ref{fig:fit_mc_tof}, \ref{fig:fit_mc_naf} and \ref{fig:fit_mc_agl}. The obtained mass resolution is shown in figure \ref{fig:massres} as a function of the kinetic energy per nucleon. It varies from $9\%$ in the AGL to $18\%$  at the highest ToF velocities. The steep increase of $\Delta m/m$ at the end of each range is an effect of the $\gamma^{4}$ term in equation \ref{eq:massres}, which is ultimately the limiting factor of the measurement. In the case of AMS, the complementary techniques for measuring the velocity allow for the identification of isotopes to be performed in broad energy range. While the ToF can separate  isotopes at energies as low as $0.2 \ \text{GeV/n}$, the RICH allows the measurements to reach up to 10 GeV/n. A study of the mass resolution in AMS-02 has been performed using helium isotopes \citep{HeMassRes}, showing similar behavior. However, the performance for helium isotopes is globally better due to the fact that the resolution of the velocity measurements increase with $Z$.

\section{Conclusion}

Studying single-charged isotopes in CRs could provide essential constraints on propagation processes, but the measurement of the fluxes of the isotopes is technically challenging because deuterons are one hundred times less abundant than protons. Several experiments have performed this measurement, but the range between 4 and 10 GeV/n remains uncharted. Due to the precise measurements of the velocity, provided especially by its Cherenkov detector, AMS-02 is able to identify isotopes through their masses in this unexplored range. In this article, we have presented and discussed a method for identifying and counting the hydrogen isotopes in ~AMS-02 data. Parametric models of the inverse mass motivated by the physical processes that happen in each of the subdetectors were built. The behavior of such parametrization as a function of velocity was discussed, also presenting how to obtain the different parameters and their uncertainties in different velocities. The fragmentation components were also included in the model. In particular, the fragmentation of helium nuclei was studied with AMS-02 data, which allowed for the construction of templates independently from interaction models present in simulation packages. The use of nuisance parameters in the fit avoided the necessity of tuning the MC simulations and allowed for a direct estimation of systematic effects coming from the knowledge of the parameters. Moreover, the proton model was readily scaled to tritons and deuterons by using a single scaling factor, allowing for the description of the different species without increasing the number of parameters substantially, hinting that such a model could also be used for the isotopic identification of $Z \geq 2$ particles as well. Finally, the quality of the fits to data was excellent and allowed for the assessment of the detector mass resolution, which is between $9$ and $18\%$ in the studied range, showing that the complementary velocity measurements provided by the ToF and RICH detectors allow for the identification of the hydrogen isotopes in the kinetic energy per nucleon range between 0.2 and 10 GeV/n.

\section*{CRediT authorship contribution statement}
\textbf{E. F. Bueno:} Conceptualization, Methodology, Software, Data Curation, Formal analysis, Visualization, Writing - Original draft preparation. \textbf{F. Barao:} Conceptualization, Methodology, Software, Writing - Review $\&$ Editing. \textbf{M. Vecchi:} Writing - Review $\&$ Editing,  Funding acquisition, Visualization, Supervision.

\section*{Declaration of competing interest}
The authors declare that they have no known competing financial interests or personal relationships that could have appeared to influence the work reported in this paper.

\section*{Acknowledgements}
This publication is part of the project "Statistical methods applied to cosmic ray anti-deuteron searches with the AMS-02 experiment" with project number 040.11.723 of the research programme Bezoekersbeurs 2019 VW, which is financed by the Dutch Research Council (NWO). We are grateful to Portuguese FCT for the financial support through grant CERN/FIS-PAR/0007/2021. 
EFB is grateful to Prof. Adriaan van den Berg for the helpful discussions during the development of this project and for his feedback on this manuscript.

\bibliography{references}

\begin{thebibliography}{10}
\expandafter\ifx\csname url\endcsname\relax
  \def\url#1{\texttt{#1}}\fi
\expandafter\ifx\csname urlprefix\endcsname\relax\def\urlprefix{URL }\fi
\expandafter\ifx\csname href\endcsname\relax
  \def\href#1#2{#2} \def\path#1{#1}\fi

\bibitem{gaisser}
T.~K. Gaisser, R.~Engel, E.~Resconi, Cosmic Rays and Particle Physics, 2nd
  Edition, Cambridge University Press, 2016.
\newblock \href {https://doi.org/10.1017/CBO9781139192194}
  {\path{doi:10.1017/CBO9781139192194}}.

\bibitem{coste}
B.~Coste, et~al.,
  \href{https://doi.org/10.1051/0004-6361/201117927}{Constraining galactic
  cosmic-ray parameters with $ z \leq 2$ nuclei}, A\&A 539 (2012) A88.
\newblock \href {https://doi.org/10.1051/0004-6361/201117927}
  {\path{doi:10.1051/0004-6361/201117927}}.
\newline\urlprefix\url{https://doi.org/10.1051/0004-6361/201117927}

\bibitem{ppchain}
E.~G. Adelberger, et~al.,
  \href{https://link.aps.org/doi/10.1103/RevModPhys.83.195}{Solar fusion cross
  sections. ii. the $pp$ chain and cno cycles}, Rev. Mod. Phys. 83 (2011)
  195--245.
\newblock \href {https://doi.org/10.1103/RevModPhys.83.195}
  {\path{doi:10.1103/RevModPhys.83.195}}.
\newline\urlprefix\url{https://link.aps.org/doi/10.1103/RevModPhys.83.195}

\bibitem{IMP3}
C.~Y. Fan, et~al.,
  \href{https://link.aps.org/doi/10.1103/PhysRevLett.17.329}{Galactic deuterium
  and its energy spectrum above 20 mev per nucleon}, Phys. Rev. Lett. 17 (1966)
  329--333.
\newblock \href {https://doi.org/10.1103/PhysRevLett.17.329}
  {\path{doi:10.1103/PhysRevLett.17.329}}.
\newline\urlprefix\url{https://link.aps.org/doi/10.1103/PhysRevLett.17.329}

\bibitem{PAMELA}
O.~Adriani, et~al.,
  \href{https://doi.org/10.3847/0004-637x/818/1/68}{Measurements of cosmic-ray
  hydrogen and helium isotopes with the {PAMELA} experiment} 818~(1) (2016) 68.
\newblock \href {https://doi.org/10.3847/0004-637x/818/1/68}
  {\path{doi:10.3847/0004-637x/818/1/68}}.
\newline\urlprefix\url{https://doi.org/10.3847/0004-637x/818/1/68}

\bibitem{BESS00}
K.~Kim, et~al.,
  \href{https://www.sciencedirect.com/science/article/pii/S0273117712000567}{Cosmic
  ray 2h/1h ratio measured from bess in 2000 during solar maximum}, Advances in
  Space Research 51~(2) (2013) 234--237, the Origins of Cosmic Rays: Resolving
  Hess's Century-Old Puzzle.
\newblock \href {https://doi.org/https://doi.org/10.1016/j.asr.2012.01.015}
  {\path{doi:https://doi.org/10.1016/j.asr.2012.01.015}}.
\newline\urlprefix\url{https://www.sciencedirect.com/science/article/pii/S0273117712000567}

\bibitem{CAPRICE98}
P.~Papini, et~al., {High-energy deuteron measurement with the CAPRICE98
  experiment}, Astrophys. J. 615 (2004) 259--274.
\newblock \href {https://doi.org/10.1086/424027} {\path{doi:10.1086/424027}}.

\bibitem{IMAX92}
G.~A. de~Nolfo, et~al.,
  \href{https://aip.scitation.org/doi/abs/10.1063/1.1324352}{A measurement of
  cosmic ray deuterium from 0.5–2.9 gev/nucleon}, AIP Conference Proceedings
  528~(1) (2000) 425--428.
\newblock \href
  {http://arxiv.org/abs/https://aip.scitation.org/doi/pdf/10.1063/1.1324352}
  {\path{arXiv:https://aip.scitation.org/doi/pdf/10.1063/1.1324352}}, \href
  {https://doi.org/10.1063/1.1324352} {\path{doi:10.1063/1.1324352}}.
\newline\urlprefix\url{https://aip.scitation.org/doi/abs/10.1063/1.1324352}

\bibitem{AMSPhysReport}
M.~Aguilar, et~al.,
  \href{https://www.sciencedirect.com/science/article/pii/S0370157320303434}{The
  alpha magnetic spectrometer (ams) on the international space station: Part ii
  — results from the first seven years}, Physics Reports 894 (2021) 1--116,
  the Alpha Magnetic Spectrometer (AMS) on the International Space Station:
  Part II - Results from the First Seven Years.
\newblock \href {https://doi.org/https://doi.org/10.1016/j.physrep.2020.09.003}
  {\path{doi:https://doi.org/10.1016/j.physrep.2020.09.003}}.
\newline\urlprefix\url{https://www.sciencedirect.com/science/article/pii/S0370157320303434}

\bibitem{amsscheme}
A.~Kounine, The alpha magnetic spectrometer on the international space station,
  International Journal of Modern Physics E 21~(8) (2012) 1230005.

\bibitem{TOFPerformance}
V.~Bindi, et~al.,
  \href{https://www.sciencedirect.com/science/article/pii/S0168900214000102}{Calibration
  and performance of the ams-02 time of flight detector in space}, Nuclear
  Instruments and Methods in Physics Research Section A: Accelerators,
  Spectrometers, Detectors and Associated Equipment 743 (2014) 22--29.
\newblock \href {https://doi.org/https://doi.org/10.1016/j.nima.2014.01.002}
  {\path{doi:https://doi.org/10.1016/j.nima.2014.01.002}}.
\newline\urlprefix\url{https://www.sciencedirect.com/science/article/pii/S0168900214000102}

\bibitem{amsrich}
F.~Giovacchini, J.~Casaus, A.~Oliva,
  \href{https://doi.org/10.1016/j.nima.2019.01.024}{The {AMS}-02 {RICH}
  detector: Status and physics results}, Nuclear Instruments and Methods in
  Physics Research Section A: Accelerators, Spectrometers, Detectors and
  Associated Equipment 952 (2020) 161797.
\newblock \href {https://doi.org/10.1016/j.nima.2019.01.024}
  {\path{doi:10.1016/j.nima.2019.01.024}}.
\newline\urlprefix\url{https://doi.org/10.1016/j.nima.2019.01.024}

\bibitem{igrf}
C.~C. Finlay, et~al.,
  \href{https://doi.org/10.1111/j.1365-246x.2010.04804.x}{International
  geomagnetic reference field: the eleventh generation}, Geophysical Journal
  International 183~(3) (2010) 1216--1230.
\newblock \href {https://doi.org/10.1111/j.1365-246x.2010.04804.x}
  {\path{doi:10.1111/j.1365-246x.2010.04804.x}}.
\newline\urlprefix\url{https://doi.org/10.1111/j.1365-246x.2010.04804.x}

\bibitem{GEANT4}
J.~Allison, et~al.,
  \href{https://www.sciencedirect.com/science/article/pii/S0168900216306957}{Recent
  developments in geant4}, Nuclear Instruments and Methods in Physics Research
  Section A: Accelerators, Spectrometers, Detectors and Associated Equipment
  835 (2016) 186--225.
\newblock \href {https://doi.org/https://doi.org/10.1016/j.nima.2016.06.125}
  {\path{doi:https://doi.org/10.1016/j.nima.2016.06.125}}.
\newline\urlprefix\url{https://www.sciencedirect.com/science/article/pii/S0168900216306957}

\bibitem{ams-helium-isotopes}
M.~Aguilar, et~al.,
  \href{https://link.aps.org/doi/10.1103/PhysRevLett.123.181102}{Properties of
  cosmic helium isotopes measured by the alpha magnetic spectrometer}, Phys.
  Rev. Lett. 123 (2019) 181102.
\newblock \href {https://doi.org/10.1103/PhysRevLett.123.181102}
  {\path{doi:10.1103/PhysRevLett.123.181102}}.
\newline\urlprefix\url{https://link.aps.org/doi/10.1103/PhysRevLett.123.181102}

\bibitem{qiyan}
Q.~Yan, V.~Choutko, A.~Oliva, M.~Paniccia,
  \href{https://www.sciencedirect.com/science/article/pii/S0375947420300221}{Measurements
  of nuclear interaction cross sections with the alpha magnetic spectrometer on
  the international space station}, Nuclear Physics A 996 (2020) 121712.
\newblock \href
  {https://doi.org/https://doi.org/10.1016/j.nuclphysa.2020.121712}
  {\path{doi:https://doi.org/10.1016/j.nuclphysa.2020.121712}}.
\newline\urlprefix\url{https://www.sciencedirect.com/science/article/pii/S0375947420300221}

\bibitem{HeFrag}
A.~U. Abdurakhimov, et~al., {Projectile fragmentation processes in He-4 nucleus
  interations at 4.5 GeV/c per incident nucleon} (8 1979).

\bibitem{HeFragNoguchi}
M.~Noguchi, et~al.,
  \href{https://www.sciencedirect.com/science/article/pii/088328899190165W}{Production
  cross sections of tritium in high energy nuclear reactions with 12 gev
  protons}, International Journal of Radiation Applications and
  Instrumentation. Part A. Applied Radiation and Isotopes 42~(6) (1991)
  577--582.
\newblock \href {https://doi.org/https://doi.org/10.1016/0883-2889(91)90165-W}
  {\path{doi:https://doi.org/10.1016/0883-2889(91)90165-W}}.
\newline\urlprefix\url{https://www.sciencedirect.com/science/article/pii/088328899190165W}

\bibitem{yijia_charge}
Y.~Jia, et~al.,
  \href{https://www.sciencedirect.com/science/article/pii/S0168900220305659}{Nuclei
  charge measurement by the alpha magnetic spectrometer silicon tracker},
  Nuclear Instruments and Methods in Physics Research Section A: Accelerators,
  Spectrometers, Detectors and Associated Equipment 972 (2020) 164169.
\newblock \href {https://doi.org/https://doi.org/10.1016/j.nima.2020.164169}
  {\path{doi:https://doi.org/10.1016/j.nima.2020.164169}}.
\newline\urlprefix\url{https://www.sciencedirect.com/science/article/pii/S0168900220305659}

\bibitem{HeMassRes}
X.~Xia, A.~Oliva,
  \href{https://www.sciencedirect.com/science/article/pii/S0168900217307167}{Mass
  resolution for helium isotopes in ams-02}, Nuclear Instruments and Methods in
  Physics Research Section A: Accelerators, Spectrometers, Detectors and
  Associated Equipment 868 (2017) 139--141.
\newblock \href {https://doi.org/https://doi.org/10.1016/j.nima.2017.07.005}
  {\path{doi:https://doi.org/10.1016/j.nima.2017.07.005}}.
\newline\urlprefix\url{https://www.sciencedirect.com/science/article/pii/S0168900217307167}

\end{thebibliography}

\end{document}